\documentclass[referee]{raa}            

\usepackage{graphicx,times}             
\usepackage{natbib}
\usepackage{amssymb,amsmath}
\usepackage{pdflscape}
\usepackage{color}
\bibpunct{(}{)}{;}{a}{}{,}

\usepackage[a4paper=true,pagebackref=true]{hyperref}
\hypersetup{colorlinks = true, linkcolor = green, anchorcolor = red, citecolor = blue, filecolor = red, pagecolor = red, urlcolor = red}

\begin{document}


   \title{Searching for AGN and Pulsar Candidates in 4FGL Unassociated Sources Using Machine Learning
}

   \volnopage{Vol.0 (20xx) No.0, 000--000}      
   \setcounter{page}{1}          

   \author{K. R. Zhu
      \inst{1}
   \and S. J. Kang
      \inst{2*}
   \and Y. G. Zheng
      \inst{1*}
   }

   \institute{Department of Physics, Yunnan Normal University, Kunming, Yunnan, 650092, People's Republic of China; {\it ynzyg@ynu.edu.cn}\\
        \and
             School of Physics and Electrical Engineering, Liupanshui Normal University, Liupanshui, Guizhou, 553004, People's Republic of China; {\it kangshiju@alumni.hust.edu.cn}\\
   \vs\no
      {\small Received~~20xx month day; accepted~~20xx~~month day}}

\abstract{ In the fourth \emph{Fermi} Large Area Telescope source catalog (4FGL), 5064 $\gamma$-ray sources are reported, including 3207 active galactic nuclei (AGNs), 239 pulsars, 1336 unassociated sources, 92 sources with weak association with blazar at low Galactic latitude and 190 other sources. We employ two different supervised machine learning classifiers, combined with the direct observation parameters given by the 4FGL fits table, to search for sources potentially classified as AGNs and pulsars in the 1336 unassociated sources. In order to reduce the error caused by the large difference in the sizes of samples, we divide the classification process into two separate steps in order to identify the AGNs and the pulsars. First, we select the identified AGNs from all of the samples, and then select the identified pulsars from the remaining.  Using the 4FGL sources associated or identified as AGNs, pulsars, and other sources with the features selected through the K-S test and the random forest (RF) feature importance measurement, we trained, optimized, and tested our classifier models. Then, the models are applied to classify the 1336 unassociated sources. According to the calculation results of the two classifiers, we show the sensitivity, specificity, accuracy in each step, and the class of unassociated sources given by each classifier. The accuracy obtained in the first step is approximately $95\%$; in the second step, the obtained overall accuracy is approximately $80\%$. Combining the results of the two classifiers, we predict that there are 583 AGN-type candidates, 115 pulsar-type candidates, 154 other types of $\gamma$-ray candidates, and 484 of uncertain types.
\keywords{gamma rays: galaxies - galaxies: active - methods: statistical}
}

   \authorrunning{Kerui Zhu et al}            
   \titlerunning{Searching for AGN and pulsar candidates in 4FGL}  

   \maketitle

\section{Introduction}\label{sec:intro}

Both the Celestial Observation Satellite (COS-B) $\gamma$-ray source catalogs (e.g., \citealt{1981RSPTA.301..519H,1987ICRC....1...88P}) and the Compton Gamma Ray Observatory (CGRO) $\gamma$-ray source catalogs (e.g., \citealt{1994ApJS...94..551F,1995ApJS..101..259T,1999ApJS..123...79H}) contain a small number of sources, most of which are unassociated sources. The identification of MeV-GeV $\gamma$-ray sources, over a long period of time, suffers from few detectors and limited angular resolution. In recent years, approximately 20 types of $\gamma$-ray sources have been identified \citep{2020ApJS..247...33A}. Most of the identified sources belong to the active galactic nuclei (AGN) category. It is commonly believed that there is a supermassive black hole (SMBH) in the centre of an AGN. Their continuum emission is characterized by high brightness and non-stellar origin. Their broad spectral energy distribution extends from radio to high-energy $\gamma$-ray bands \citep{2019arXiv190106507K}. In the widely accepted unified model paradigm \citep{1995PASP..107..803U,1997ARA&A..35..445U}, an AGN is usually associated with a jet that originates from the central SMBH and is filled with relativistic plasmas. Due to their extreme characteristics, the jet of an AGN is an ideal object for studying on the acceleration mechanism of high-energy particles. In addition, pulsars are another major observed type; the pulsars' high energy emission mechanism is an open issue. Considering the different locations of the emission region \citep{1998ApJ...508..328H}, either the polar cap model \citep{1998ASPC..138..281R,1998nspt.conf..311H} or the outer gap model \citep{1986ApJ...300..500C,1996ApJ...470..469R,2014Sci...344..159R} is used to interpret the high-energy emission of pulsars. The latter model is more popular \citep{2016ApJ...820....8S} since a large number of radio-quiet $\gamma$-ray pulsars have been identified by the fermi-LAT \citep{2009Sci...325..840A,2010ApJ...725..571S}. However, additional evidence is still required.

In 2008, a new era in the classification of observations began to emerge. High-energy observations have been included in the Fermi catalogs; an abundance of $\gamma$-ray sources have been discovered. Over the last decade, the Fermi-LAT source catalog (FGL) has evolved, including the regular releases of the 0FGL (3 months, \citealt{2009ApJS..183...46A}), 1FGL (11 months, \citealt{2010ApJS..188..405A}), 2FGL (2 years, \citealt{2012ApJS..199...31N}), and 3FGL (4 years, \citealt{2015ApJS..218...23A}). Neglecting the incomplete 0FGL, the 1FGL contains 1451 sources including 630 unassociated sources \citep{2010ApJS..188..405A}. Then the 2FGL reduces the number of these unassociated sources to 576; this catalog contains a total of 1873 sources. The 3FGL contains 3033 sources of which approximately one third are unassociated \citep{2015ApJS..218...23A}. Recently, the Fermi-LAT collaboration has provided a release of the fourth Fermi-LAT source catalog (4FGL)\footnote{\url{https://fermi.gsfc.nasa.gov/ssc/data/access/lat/8yr_catalog/}}. This catalog exhibits the new $\gamma$-ray observation results of an eight-year period from 2008 to 2016 in the 50 MeV to 1 TeV energy range with 4$\sigma$ confidence level. The 5064 sources contained in the 4FGL are divided into 23 categories \citep[see][]{2020ApJS..247...33A}, in which the number of sources of eight classes of AGNs is 3207, accounting for $63.3\%$ of the total sources. Besides, 239 sources are pulsars, 1336 sources are unassociated, 191 sources are identified in 11 other categories (i.e., pulsar wind nebula and normal galaxy, etc), and 92 sources labeled as ``UNK/unk" in the 4FGL table, which are the sources with weak association with blazar at low Galactic latitude (marked as UNK in the work). Since the AGNs and pulsars are important for the field of high-energy astrophysics, we evaluate the potential classification of unassociated sources and confirm the AGN and pulsar candidates for the expanding samples.

Machine learning (ML) techniques have become more popular in the field of data mining and data analysis and are receiving attention in a wide variety of domains, including the analysis of astronomical databases \citep{Ball2010DATA,2012MNRAS.424L..64M,2010SPIE.7740E..0LP,2014ApJ...782...41D,2016MNRAS.462.3180C,2016ApJ...820....8S,
2017ICRC...35..600L,2017MNRAS.470.1291S,2019arXiv190407248,2019ApJ...872..189K,2019ApJ...887..134K,2019MNRAS.486.3415L,2019ApJ...881L...9F,
2019arXiv191202934F}. As a cutting-edge multiple-subjects-crossing subject, ML is divided into supervised machine learning (SML) and unsupervised machine learning (USML) algorithms. Based on the clustering algorithm, the USML is utilized to identify the potentially complex relationships among samples. Alternatively, if we focus primarily on the labels of datasets provided artificially, we can employ SML algorithms to realize the classification and regression \citep{2019arXiv190407248}. The aim of SML classifiers is to establish judgment criteria based on known samples to predict the classification of unknown samples. A wide variety of SML algorithms are available, including logistic regression, decision trees, random forest, support vector machines, neural networks, Bayesian networks, Gaussian finite mixture models, artificial neural network, and many others (e.g., see \citealt{2012msma.book.....F}; \citealt{Kabacoff2015R}).

In recent years, ML algorithms have been widely used in Fermi data analysis. Many investigators have utilized them to explore the nature of unidentified $\gamma$-ray sources. For example, searching for AGNs \citep{2012MNRAS.424L..64M,2014ApJ...782...41D,2016ApJ...820....8S} and pulsars \citep{2012MNRAS.424L..64M,2016ApJ...820....8S,2020MNRAS.tmp..163L} in unassociated sources, or evaluating the optical classification of Fermi blazar candidates of uncertain type (BCUs) \citep{2013MNRAS.428..220H,2016MNRAS.462.3180C,2017ICRC...35..600L,2017MNRAS.470.1291S,2019ApJ...872..189K,2019ApJ...887..134K,
2019MNRAS.486.3415L}.

In the present context, we employ two SML classification methods of both random forest (RF) and artificial neural network (ANN) to evaluate the potential classification of the 1336 unassociated sample sources in the 4FGL catalog. The aim is to obtain more potential AGN, pulsar, and other $\gamma$-ray sources (non-AGN and non-pulsar) candidates. The remainder of this paper is organized as follows. In Section 2, we describe the dataset from the 4FGL and select features using K-S test and RF feature importance measurement. In Section 3, we review SML classification algorithms, dataset partitioning and normalization, the creation and validation of two individual algorithms (RF and ANN). In Section 4, we test the individual algorithms and composition algorithm, then, apply the composition model to the 1336 unassociated sources. Some discussions and the conclusion are given in Section 5.

\section{Dataset preparation} \label{sec:dataset}

In the new release of the 4FGL catalog fits table\footnote{\url{https://fermi.gsfc.nasa.gov/ssc/data/access/lat/8yr_catalog/gll_psc_v21.fit}}, 5064 $\gamma$-ray sources above a 4$\sigma$ confidence level are reported, and these are divided into 23 categories. Nevertheless, not all samples are available.
The nature of UNKs has not been defined, though there is a weak association between UNKs and blazar candidates. Moreover, the bright background at the low Galactic latitude impact the observation of UNKs, which may lead to the deviation of the classification process. So, the 92 UNK sources are removed. In the classification, eight classes of AGNs, such as flat spectrum radio quasars, BL Lac objects, and Seyfert galaxy, are labeled as agn. Similarly, we label the pulsars as psr, unassociated sources as unass, and the rest of the sources that are identified as other $\gamma$-ray sources are labeled as other. The details of the 4972 sources that belong to different categories or labels are shown in Table \ref{Tab1}.

As seen in Table \ref{Tab1}, the sample is unbalanced. More specifically, the number of AGNs is approximately 15 times as the number of pulsars or other types, which can significantly affect the classification results. In order to reduce the influence of the imbalances and improve the prediction accuracy, we divide the classification process into two steps. Firstly, we select the AGN candidates in all of the unassociated samples, and then select the pulsar candidates in the remaining non-AGN samples for the last step. In this way, we expand the non-AGN samples and reduce the error. The classification is done step by step; thus, there are distinct datasets in the two steps. (see Table \ref{Tab2}).

\begin{table*}
\centering
\caption{The label of 4FGL samples} \label{Tab1}
\begin{tabular}{cccccccc} 
\hline\hline
Description	&~&	Designator&~&{Source count}	&~&Label\\
\hline
Non-blazar active galaxy	               &&agn                      &&11	          &&agn\\
Blazar candidate of uncertain type	       &&bcu                      &&1312	      &&agn\\
FSRQ type of blazar	                       &&fsrq                     &&694	          &&agn\\
Compact Steep Spectrum radio source	       &&css                      &&5	          &&agn\\
Narrow line Seyfert 1	                   &&nlsy1                    &&9	          &&agn\\
Radio galaxy	                           &&rdg                      &&42	          &&agn\\
Seyfert galaxy	                           &&sey                      &&1	          &&agn\\
Steep spectrum radio quasar	               &&ssrq                     &&2	          &&agn\\
BL Lac type of blazar	                   &&bll                      &&1131	      &&agn\\
Binary	                                   &&bin                      &&1	          &&other\\
Normal galaxy (or part)	                   &&gal                      &&3	          &&other\\
Globular cluster	                       &&glc                      &&30	          &&other\\
High-mass binary	                       &&hmb                      &&8	          &&other\\
Low-mass binary	                           &&lmb                      &&2	          &&other\\
Nova	                                   &&nov                      &&1	          &&other\\
Pulsar wind nebula	                       &&pwn                      &&17	          &&other\\
Starburst galaxy	                       &&sbg                      &&7	          &&other\\
Star-forming region	                       &&sfr                      &&3	          &&other\\
Supernova remnant	                       &&snp                      &&40	          &&other\\
Supernova remnant / Pulsar wind nebula	   &&spp                      &&78	          &&other\\
Pulsar	                                   &&psr                      &&239	          &&psr\\
Unassociated	                           &&{}                       &&1336	      &&unass\\
\hline
\end{tabular}\\
{Note: Column 1: Descriptions of sources for different classes \citep{2020ApJS..247...33A}.
Column 2: Designator of sources for different classes.
Column 3: Sources court of different classes.
Column 4: The label of different sources used in this paper.}
\end{table*}

Each source in the 4FGL catalog contains 333 columns of observed data \citep{2020ApJS..247...33A}. Excluding strings, missing columns, columns without physical significance, errors, and historical data, there are 36 usable features. [$F_1-F_7$]: integral photon flux in the band of 50 to 100 MeV, 100 to 300 MeV, 300 MeV to 1 GeV, 1 to 3 GeV, 3 to 10 GeV, 10 to 30 GeV and 30 to 300 GeV, respectively; [$\nu F_{\nu1}-\nu F_{\nu7}$]: spectral energy distribution over the seven bands; [$GLON$/$GLAT$]: galactic longitude/latitude; [$E_{100}$]: energy flux from 100 MeV to 100 GeV; [$F_{1000}$]: integral photon flux from 1 to 100 GeV; [$Signif\_ Avg$]: source significance in $\sigma$ units over the 100 MeV to 1 TeV band; [$E_{Pivot}$]: the energy at which error on differential flux is minimal; [$K_{PL}$, $PL\_ Index$]: differential flux at pivot energy, photon index in PL (powerlaw) fit; [$K_{LP}$, $LP\_ Index$,$LP\_ beta$]: differential flux, photon index at pivot energy, curvature in LP (logarithmic parabola) fit; [$K_{PLEC}$, $PLEC\_ Index$,$PLEC\_ Expfactor$ and $PLEC\_Exp\_Index$]: differential flux at pivot energy, low-energy photon index, exponential factor and index in PLEC (powerlaw with superexponential cutoff) fit; [$LP\_ SigCurv$/$PLEC\_ SigCurv$]: significance of the fit improvement between PL and LP/PLEC in $\sigma$ units; [$Npred$]: predicted number of events in the model; [$Variability\_ Index$]: variability index over the full catalog interval; [$Variability2\_ Index$]: variability index over two-month intervals; [$Frac\_Variability$/$Frac2\_Variability$]: fractional variability computed from the fluxes in each year/two months.

In order to facilitate normalization and reduce the computational demands of subsequent steps in the process, we calculate the logarithm of the higher scale features (flux, energy. etc).

Since different features play different roles in the classifiers, the selection of suitable input features for the SML is necessary. Noticing that, i) More input features do not always result in higher accuracy \citep{2019ApJ...887..134K}; ii) More features need more computation; iii) The favorable features for the selection of the AGNs are different from those for pulsars, we further select the features for the two steps from the 36 usable features.

The Kolmogorov-Smirnov test (K-S test) is a two-sample hypothesis test method, which is often used to evaluate the significance of the distribution difference of the same measurement in two samples (e.g., \citealt{2014MNRAS.441.3375X,2020ApJ...891...87K}). In particular, the K-S test can also be used for feature selection \citep[e.g.,][]{2019ApJ...872..189K,2019ApJ...887..134K}, based on the principle that the greater the distribution difference of the two samples over a feature, the more favorable the feature is in SML classifiers. In addition, feature importance provides a metric on the feature performance evaluation in the RF algorithm. Here, this is measured using the function ``\emph{importance}'' from the package ``\emph{randomForest}'' \citep{randomForest}. In summary, these two test methods are employed to evaluate the 36 usable features. For the purpose of seasoning with the two-step classification process, we first test the features of AGNs and non-AGNs; then, the same process is applied between pulsars and other $\gamma$-ray sources. The pulsars and other $\gamma$-ray sources are labelled as non-AGN in the first step.

\begin{table*}
\centering
\caption{Results of test}\label{Tab2}
\resizebox{\textwidth}{!}{
\begin{tabular}{lcccclccc}
\hline\hline
\multicolumn{4}{c}{{First step}}   & &\multicolumn{4}{c}{Second step}      \\
\cline{1-4} \cline{6-9}	 			
feature&{$D$} & {$p$}&{$RF\ Gini$}& &feature&{$D$} & {$p$}&{$RF\ Gini$} \\
\hline
$logF_4$	&	0.605	&	0	&	19.13	&	&	$PLEC\_SigCurv$	&	0.547	&	0	&	17.62	\\
$log\nu F_{\nu4}$	&	0.603	&	0	&	19.86	&	&	$LP\_SigCurv$	&	0.518	&	0	&	16.28	\\
$LP\_SigCurv$	&	0.598	&	0	&	17.40	&	&	$LP\_beta$	&	0.434	&	0	&	13.17	\\
$PLEC\_SigCurv$	&	0.591	&	0	&	17.69	&	&	$PLEC\_Expfactor$     	&	0.399	&	$<1\times10^{-6}$ 	&	12.78	\\
$PLEC\_Expfactor$	&	0.589	&	0	&	18.93	&	&	$Signif\_Avg$	&	0.394	&	$<1\times10^{-6}$ 	&	16.47	\\
$logF_1000$	&	0.588	&	0	&	19.42	&	&	$log\nu F_{\nu7}$	&	0.379	&	$<1\times10^{-6}$ 	&	14.98	\\
$Frac\_Variability$	&	0.560	&	0	&	17.21	&	&	$PLEC\_Index$	&	0.375	&	$<1\times10^{-6}$ 	&	10.63	\\
$LP\_beta$	&	0.555	&	0	&	19.29	&	&	$logF_7$	&	0.350	&	$<1\times10^{-6}$ 	&	12.18	\\
\cline{6-9}
$log\nu F_{\nu3}$	&	0.545	&	0	&	18.23	&	&	$logK_{LP}$	&	0.281	&	$<1\times10^{-6}$ 	&	8.19	\\
$logF_5$	&	0.530	&	0	&	16.72	&	&	$logK_{PLEC}$	&	0.281	&	$<1\times10^{-6}$ 	&	6.66	\\
$logF_3$	&	0.525	&	0	&	18.57	&	&	$logK_{PL}$	&	0.267	&	$<1\times10^{-6}$ 	&	7.96	\\
$log\nu F_{\nu5}$	&	0.508	&	0	&	15.93	&	&	$E_{Pivot}$	&	0.262	&	$<1\times10^{-6}$ 	&	5.93	\\
$logE_100$	&	0.503	&	0	&	18.02	&	&	$Npred$	&	0.195	&	$6.48\times10^{-4}$	&	10.52	\\
$Variability\_Index$   	&	0.457	&	0	&	18.15	&	&	$logF_5$	&	0.186	&	$1.33\times10^{-3}$	&	8.33	\\
$Npred$	&	0.446	&	0	&	13.72	&	&	$log\nu F_{\nu6}$	&	0.181	&	$1.97\times10^{-3}$	&	8.70	\\
$PLEC\_Index$	&	0.445	&	0	&	18.48	&	&	$logF_1000$	&	0.176	&	$2.88\times10^{-3}$	&	9.34	\\
$Variability2\_Index$  	&	0.382	&	0	&	19.64	&	&	$log\nu F_{\nu4}$	&	0.172	&	$3.84\times10^{-3}$	&	8.46	\\
$Frac2\_Variability$   	&	0.371	&	0	&	19.01	&	&	$log\nu F_{\nu5}$	&	0.170	&	$4.32\times10^{-3}$	&	7.60	\\
$logK_{LP}$	&	0.360	&	0	&	12.75	&	&	$logF_4$	&	0.167	&	$5.60\times10^{-3}$	&	9.16	\\
$logK_{PLEC}$	&	0.360	&	0	&	12.75	&	&	$logF_6$	&	0.159	&	$9.59\times10^{-3}$	&	8.78	\\
\cline{1-4}
$logK_{PL}$	&	0.335	&	0	&	12.67	&	&	$GLAT$	&	0.153	&	$1.44\times10^{-2}$	&	5.04	\\
$GLAT$	&	0.329	&	0	&	9.76	&	&	$PL\_Index$	&	0.137	&	$3.83\times10^{-2}$	&	6.44	\\
$Signif\_Avg$	&	0.289	&	0	&	14.89	&	&	$Frac\_Variability$	&	0.136	&	$4.08\times10^{-2}$	&	7.44	\\
$log\nu F_{\nu2}$	&	0.288	&	0	&	14.15	&	&	$GLON$	&	0.133	&	$4.68\times10^{-2}$	&	4.51	\\
$logF_2$	&	0.272	&	0	&	12.65	&	&	$log\nu F_{\nu3}$	&	0.116	&	$1.18\times10^{-1}$	&	6.97	\\
$PL\_Index$	&	0.261	&	0	&	12.10	&	&	$logF_3$	&	0.110	&	$1.57\times10^{-1}$	&	7.58	\\
$log\nu F_{\nu7}$	&	0.195	&	$<1\times10^{-6}$	&	10.51	&	&	$Frac2\_Variability$  &		0.104	&	$2.02\times10^{-1}$	&	1.89	\\
$LP\_Index$	&	0.188	&	$<1\times10^{-6}$	&	13.80	&	&	$LP\_Index$	&	0.100	&	$2.37\times10^{-1}$	&	5.09	\\
$GLON$	&	0.187	&	$<1\times10^{-6}$	&	3.72	&	&	$logE_100$	&	0.095	&	$2.97\times10^{-1}$	&	9.07	\\
$logF_6$	&	0.169	&	$<1\times10^{-6}$	&	14.92	&	&	$logF_2$	&	0.089	&	$3.75\times10^{-1}$	&	3.31	\\
$E_{Pivot}$	&	0.164	&	$<1\times10^{-6}$	&	14.79	&	&	$log\nu F_{\nu2}$	&	0.089	&	$3.75\times10^{-1}$	&	5.54	\\
$log\nu F_{\nu6}$	&	0.155	&	$<1\times10^{-6}$	&	15.35	&	&	$Variability2\_Index$ &		0.086	&	$4.17\times10^{-1}$	&	-2.30	\\
$logF_7$	&	0.152	&	$<1\times10^{-6}$	&	9.62	&	&	$Variability\_Index$  	&	0.085	&	$4.33\times10^{-1}$	&	0.46	\\
$log\nu F_{\nu1}$	&	0.132	&	$3.46\times10^{-6}$	&	10.85	&	&	$logF_1$	&	0.072	&	$6.35\times10^{-1}$	&	4.65	\\
$logF_1$	&	0.128	&	$8.41\times10^{-6}$	&	10.85	&	&	$log\nu F_{\nu1}$	&	0.072	&	$6.35\times10^{-1}$	&	5.79	\\
$PLEC\_Exp\_Index$	&	0.009	&	1	&	0.37	&	&	$PLEC\_Exp\_Index$	&	0.013	&	1	&	1.00	\\
\hline
\end{tabular}}\\
{Note: Column 1-4 show the test results of the 36 ``usable'' features for the first step.
Columns 5-8 show the test results of the 36 ``usable'' features for the second step. Above the horizontal line are the features we selected, 20 for the first step and 8 for the second step Column 1 and 5: Tested feature name .
Columns 2-3 and 6-7: Value of test statistic ($D$) and p-value ($p$) for the two-sample K-S test.
Columns 4 and 8: RF mean decrease in accuracy factors  given by the function ``\emph{importance}'' from package ``\emph{randomForest}'' \citep{randomForest}.}
\end{table*}

The test results are shown in Table \ref{Tab2}. In the K-S test, the statistical value $D$ represents the distribution difference level of the feature in the two subclasses, while $p$ represents the probability that the feature conforms to the same distribution. The RF \emph{Gini} is the mean decrease in accuracy factors given by the measured RF feature importance; these tend to follow the same pattern as the K-S test. According to the selection criteria ($D\ge0.35$ in the K-S test), 20 better features selected in the first step and eight better features selected in the second step are shown in Table \ref{Tab2}. The features above the horizontal line are the features we selected.

\section{Establish classifier models} \label{sec:classifier}
\subsection{Classification methods }

In the field of SML, the dataset contains a certain number of objects. Each object has its own features and a target variable; for classifiers, is the target variable is also called a label \citep{2019arXiv190407248}. For our work, the dataset contains the 5065 sources from the 4FGL catalog, the features are the observation data of these sources, and the target variables are the classes of the source.

In a classifier, the model learns the corresponding relationships between input features and target variables. Then, inputting the features of the unknown samples the model outputs the probability $P$ (usually normalized to 0-1) of each sample. Based on the classification threshold (the default value is 0.5 in two-sample classifiers), the unknown samples can be divided into two classes. Therefore, the dataset is further divided according to the role it plays in the classification process. The known samples are put into the training, validation, and test datasets in a certain proportion. The training set is used to train the classification model. The validation set can help to find the best combination of hyper-parameters (parameters of classifiers, such as the number of trees in RF), classification threshold of different algorithms, or prevent over-fitting (see \citealt{2019arXiv190407248} for more details). The test set is used to evaluate the classifiers' performance in terms of accuracy, sensitivity etc.

This work employs both an RF and ANN algorithms, which contain different origins and characteristics. The RF algorithms are derived from a ``decision tree'' algorithm, which is a simple classifier algorithm (see e.g., \citealt{UTGOFF1989Incremental,Duda2001Pattern} for more details). The principle of a decision tree algorithm is to build nodes to make one-to-one judgments, and a large number of nodes constitutes a ``tree''. However, a limitation of the ``decision tree'' is an over-fitting situation, which leads to a decrease in the accuracy of judgment \citep{Duda2001Pattern}. An RF algorithm addresses the over-fitting problem by utilizing a combination of a large number of decision trees with weight consideration for each tree \citep{Breiman2001Random}. Compared with the ``decision tree'' \citep{JMLR:v15:delgado14a}, it is a more efficient and accurate classifier. Yet, a traditional RF \citep{Breiman2001Random} requires a ``clean'' dataset, which means that the input of uncertain features or missing values is unfavorable. Recently, the probabilistic random forest (PRF) algorithm has been proposed to deal with uncertain datasets \citep{2019ascl.soft03009R,2019AJ....157...16R}, which also makes the RF algorithms more suitable for astronomy data. As a mature ML classification algorithm, RF is very popular in the field of astronomical data analysis (e.g., \citealt{2003sca..book..243B,2019MNRAS.490.2367C,2019MNRAS.488.4858H,2019ApJ...872..189K,2019ApJ...887..134K}).

The ANN algorithms are based on the structure of human brain neurons, and they are used in both SML and USML. Owing to their nonlinearity, diversity characteristics, and wide applicability in the areas of regression, classification, and model prediction, the ANN algorithms are widely used in astronomy (e.g., \citealt{2004A&A...423..761V, 2010MNRAS.406..342B, 2010MNRAS.407.2443E, 2013ApJ...772..140B, 2014A&A...568A.126B, 2016MNRAS.458L..34E, 2016MNRAS.457.2086T, 2018A&A...616A..69B, 2018ApJ...858..114H, 2018NatAs...2..151N, 2018MNRAS.476.1151P, 2019MNRAS.484..294D}). The network structure is generally divided into an input layer, one or more hidden neuron layers composed of a large number of nodes, and an output layer. However, the input and output data are generally normalized, which means that normalization and de-normalization conversions are necessary. In addition, there may be extensive computational demands resulting from a large number of neurons \citep{2019arXiv190608864H}.

Currently, the R language (\citealt{R}, version for 3.5.1) is a convenient platform to implement various classifier algorithms. The RF and ANN algorithms are realized using the package ``\emph {randomforest}'' \citep{randomForest} and ``\emph{RSNNS}'' (the Stuttgart Neural Network Simulator for \emph{R} language; see, \citealt{RSNNS}), respectively.

For the purpose of evaluating the performance of classifiers, we also employed some other methods. The confusion matrix is a common metrics in classifiers test \citep{2019arXiv190407248}. The ``\emph{class$\_$eval}'' \citep{2003sca..book..243B} is a graph function that realizes the visualization of the confusion matrix. More specifically, the horizontal axis indicates the predicted label, the vertical axis represents the true label, and the accuracies appear on top of them. In addition, the function ``\emph{performance}'' \citep{Kabacoff2015R} provides a way to calculate several model performance parameters, including sensitivity (true positive rate), specificity (true negative rate), and overall accuracy based on the confusion matrix. The curves of the receiver operating characteristic (ROC) is another important classifier performance evaluation metric, which consist of points at which the sensitivity is plotted against the specificity at different classification thresholds \citep{2016ApJ...820....8S} or the true positive rate is plotted against the false positive rate \citep{2019arXiv190407248}. The $pROC$ package \citep{pROC} is employed to obtain the ROC curves for sensitivity against specificity of different classifiers and the values of the areas under the ROC curves (AUC) in this work.

In the first step of agn selection, all of the sources in the sample with 20 selected features (see Table \ref{Tab2}) are taken into account. A total of 3207 AGNs and 429 non-AGNs, containing 239 pulsars and 190 other sources, are randomly divided into training set, validation set and test set. Considering the impact of randomness on data set partitioning on a single result, we adopt a fixed randomness (random seed of ``\emph{12345}'') and uniqueness ratio (6:2:2) between training, validation and test set. Again, the 239 pulsars and 190 other gamma-ray sources would be randomly (random seed of ``\emph{12345}'') divided into training sets, validation sets, and test sets in the same ratio (6:2:2) in the second step. In order to obtain uniform results, we also set the random seed as ``\emph{12345}'' during random forest and artificial neural network training.

In addition, the normalization of input features is necessary in the artificial neural network, but not in the RF, and the input target variables (class labels) of the training, validation and test set need to be decoded into a binary matrix. For the purpose of features normalization, we call the ``\emph{normalizeData}'' function in the RSNNS package, where there are three modes to choose from, i.e., type``\emph{0$\_$1}'' (normalized to the interval from 0 to 1), type ``\emph{center}'' (the data is centered, i.e., the mean is subtracted), and type ``\emph{norm}'' (mean zero, variance one) \citep{RSNNS}. In the work, our feature normalization type is ``\emph{norm}''. In addition, the function ``\emph{decodeClassLabels}'' is adopted for decoding class labels to a binary matrix, while the function ``\emph{encodeClassLabels}'' is the approach utilized for encoding the binary matrix.

\begin{table*}{}
\centering
\caption{The best hyper-parameter combines of classifiers}\label{Tab3}
\begin{tabular}{ccccccccccc}
\hline\hline
~&~&\multicolumn{4}{c}{{RF}}&~&\multicolumn{4}{c}{{ANN}} \\
\cline{3-6}\cline{8-11}
~&~&mtry&ntree&Auc&threshold&&size&maxit&Auc&threshold\\
\hline
Step 1&~&3&102&0.992&0.809$^\star$&&9&149&0.988&0.785$^\star$\\
~&~&~&~&~&~&&9&150&0.988&0.785\\
\hline
Step 2&~&4&56&0.829&0.580$^\star$&&c(4,12)&142&0.796&0.345$^\star$\\
~&~&4&65&0.829&0.600&&~&~&~&~\\
~&~&4&78&0.829&0.390&&~&~&~&~\\
\hline
\end{tabular}\\
{Note: Column 1: Step 1 for selection of AGN and step 2 for selection of pulsar.
Columns 2-5: The best hyper-parameter combines, the corresponding Auc value and threshold of RF classifier.
Columns 6-9: The best hyper-parameter combines, the corresponding Auc value and threshold of ANN classifier.
The hyper-parameter combine marked with a symbol $\star$ is the combine obtained in present context.
}
\end{table*}

\subsection{Model Creation and validation: RF}

\begin{figure}{}
\centering
\includegraphics[height=5cm,width=7cm]{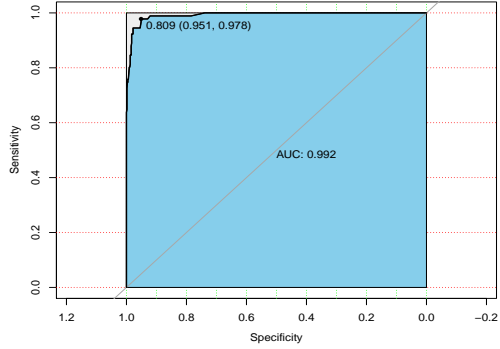}
  \caption{The ROC curves of the RF classifier with the best hyper-parameter combination for the validation set in the first step. The text in the figure are the AUC value, the optimal threshold and the corresponding sensitivity and specificity.}
 \label{fig:fig1}
\end{figure}

\begin{figure}{}
\centering
\includegraphics[height=3cm,width=4.8cm]{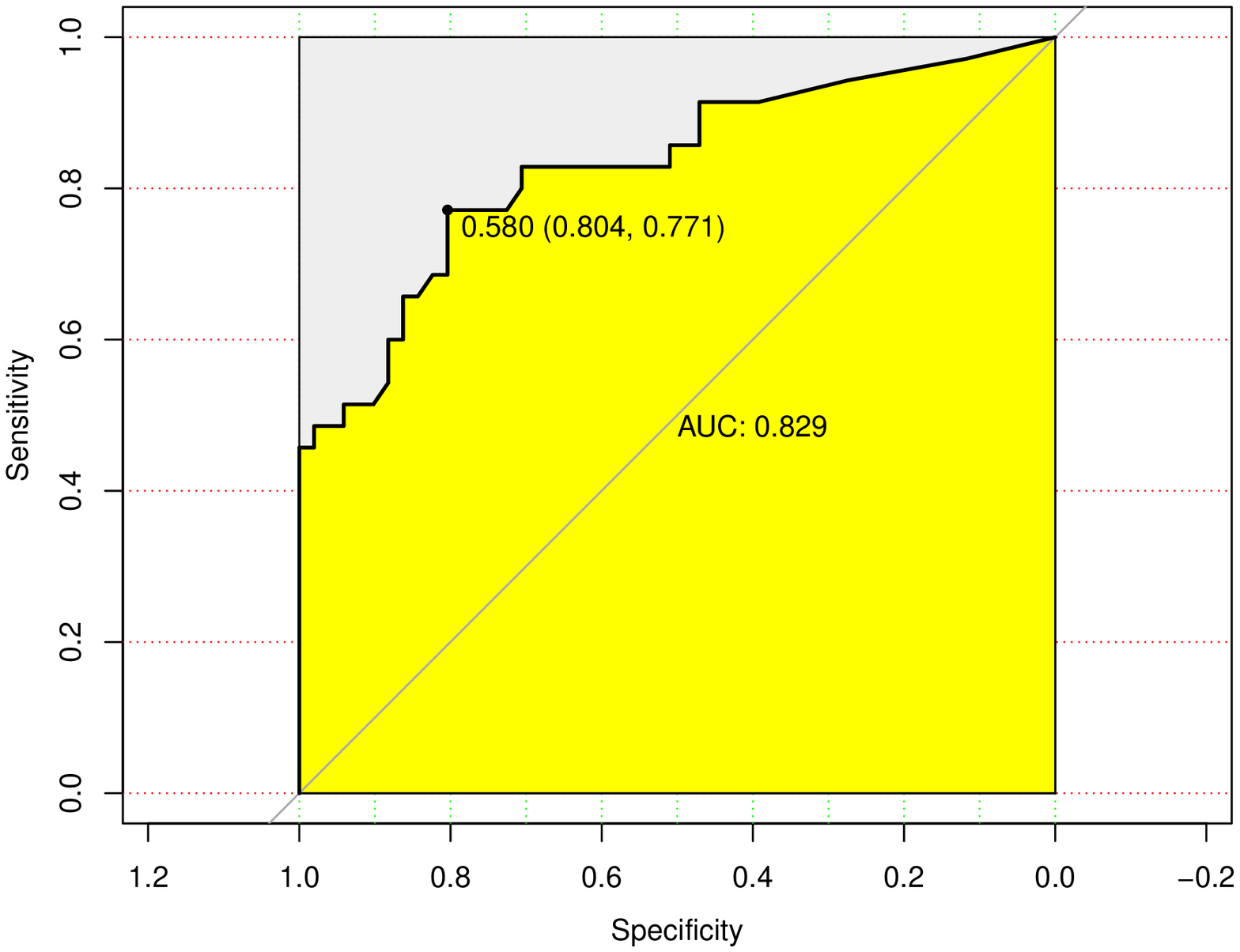}
 \includegraphics[height=3cm,width=4.8cm]{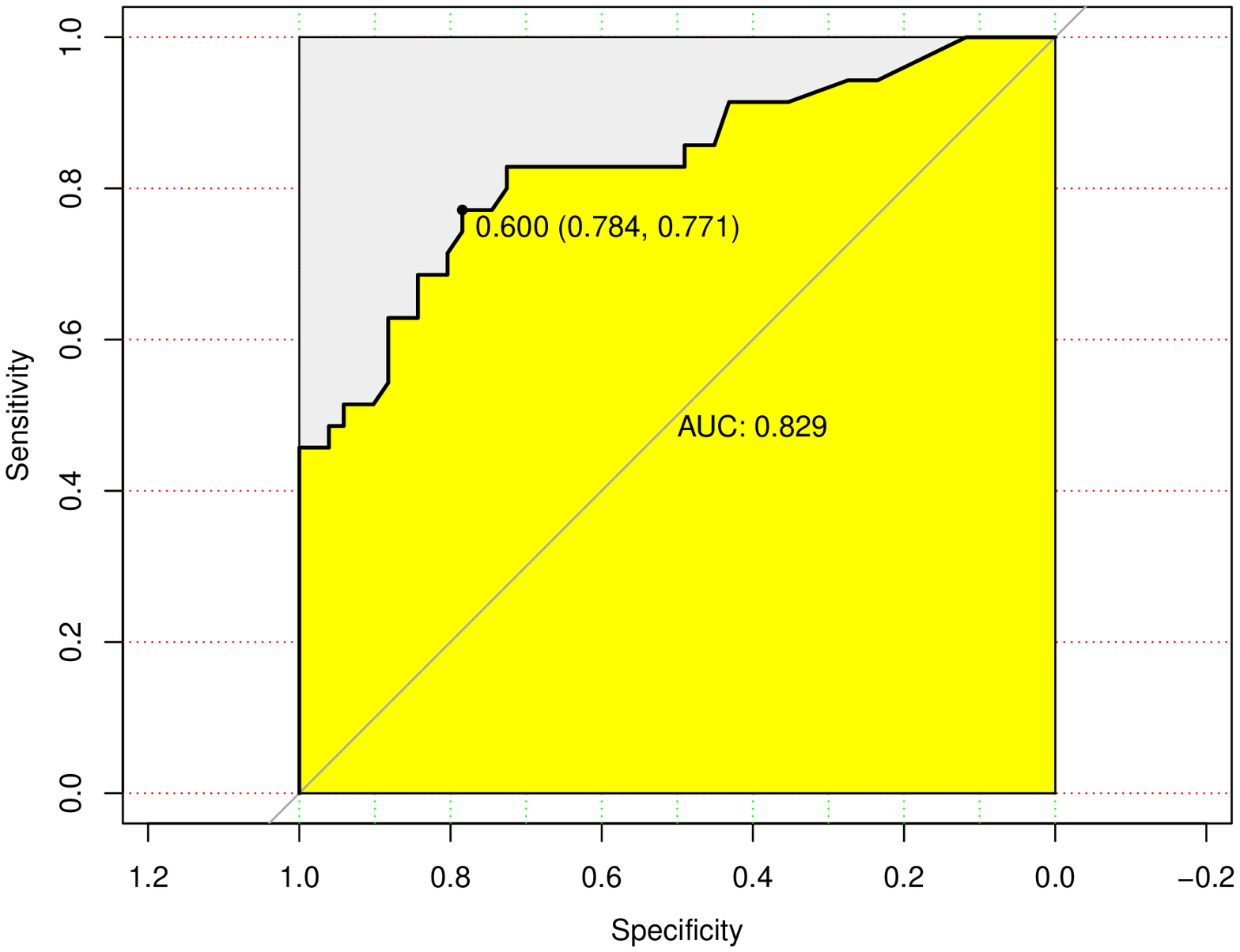}
  \includegraphics[height=3cm,width=4.8cm]{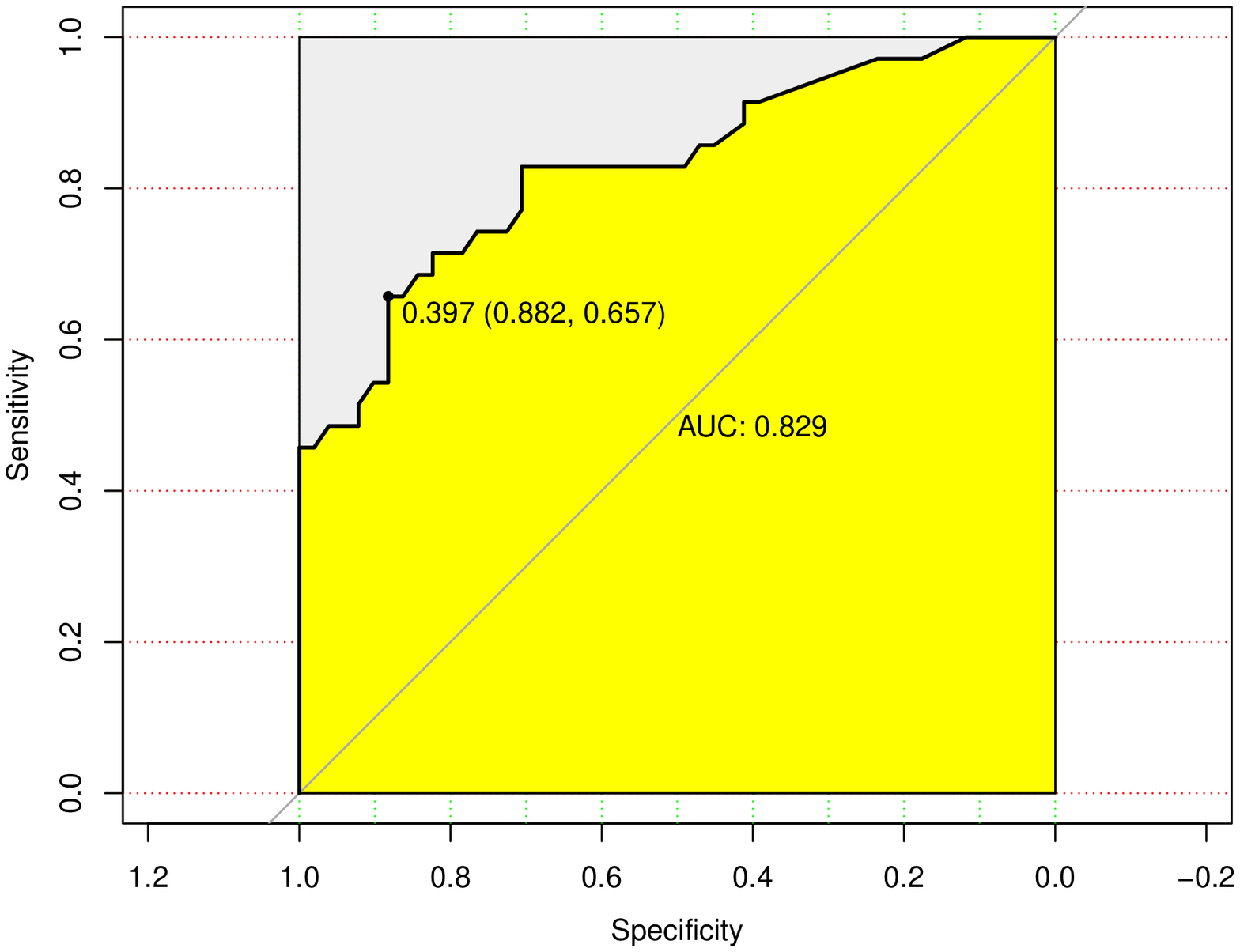}
 \caption{The ROC curves of the RF classifier with the three best hyper-parameter combinations for the validation set in the second step. The text in the figure are the AUC value, the optimal threshold and the corresponding sensitivity and specificity. The different panels correspond to different hyper-parameter combinations}
 \label{fig:fig2}
\end{figure}

 In the package ``\emph {randomforest}'' \citep{randomForest}, we build the classifier from function ``\emph {randomforest}'', which contains two hyper-parameters, ``\emph {ntree}'' and ``\emph {mtry}''. The ``\emph {ntree}'' represents the number of trees to grow, and the default value is 500. The ``\emph {mtry}'' shows the number of features randomly sampled as candidates at each split ranged from 1 to 8. For classifier, the default value is $\sqrt{n}$, where n is number of features. With all the combinations of the ``\emph {ntree}'' in the range of 50 to 750 and ``\emph {mtry}'' in the range of 1 to 8, we train the classifiers using the training set, and calculate the AUCs of the validation set of different hyper-parameter combinations. The hyper-parameter combinations corresponding to the maximum AUC value for the first step are shown in Table \ref{Tab3} and the corresponding ROC curves are shown in Figure \ref{fig:fig1}, while those for the second step are shown in  Table \ref{Tab3} and Figure \ref{fig:fig2}. In the first step, there is a best hyper-parameter combination, ``\emph {ntree=102}'' and ``\emph {mtry=3}'', respectively. The best AUC is 0.992, and the thresholds are 0.809 (see  Figure \ref{fig:fig1}). In the second step, there are three best hyper-parameter combinations, ``\emph {ntree=56}'' and ``\emph {mtry=4}'', ``\emph {ntree=65}'' and ``\emph {mtry=4}'', ``\emph {ntree=78}'' and ``\emph {mtry=4}'', respectively. The best AUC is 0.829, and the thresholds are 0.580, 0.600, 0.390, respectively (see  Figure \ref{fig:fig2}).

Accordingly, in several hyper-parameters combinations with the highest AUC values, we adopt the prior one, i.e., we set ``\emph {ntree=102}'' and ``\emph {mtry=3}'' for RF in the first step, while the threshold is set to 0.809. ``\emph {ntree=56}'', ``\emph {mtry=4}'' and the threshold is set to 0.580 in the second step.

\subsection{Model Creation and validation: ANN}

\begin{figure}
\centering
 \includegraphics[height=5cm,width=7cm]{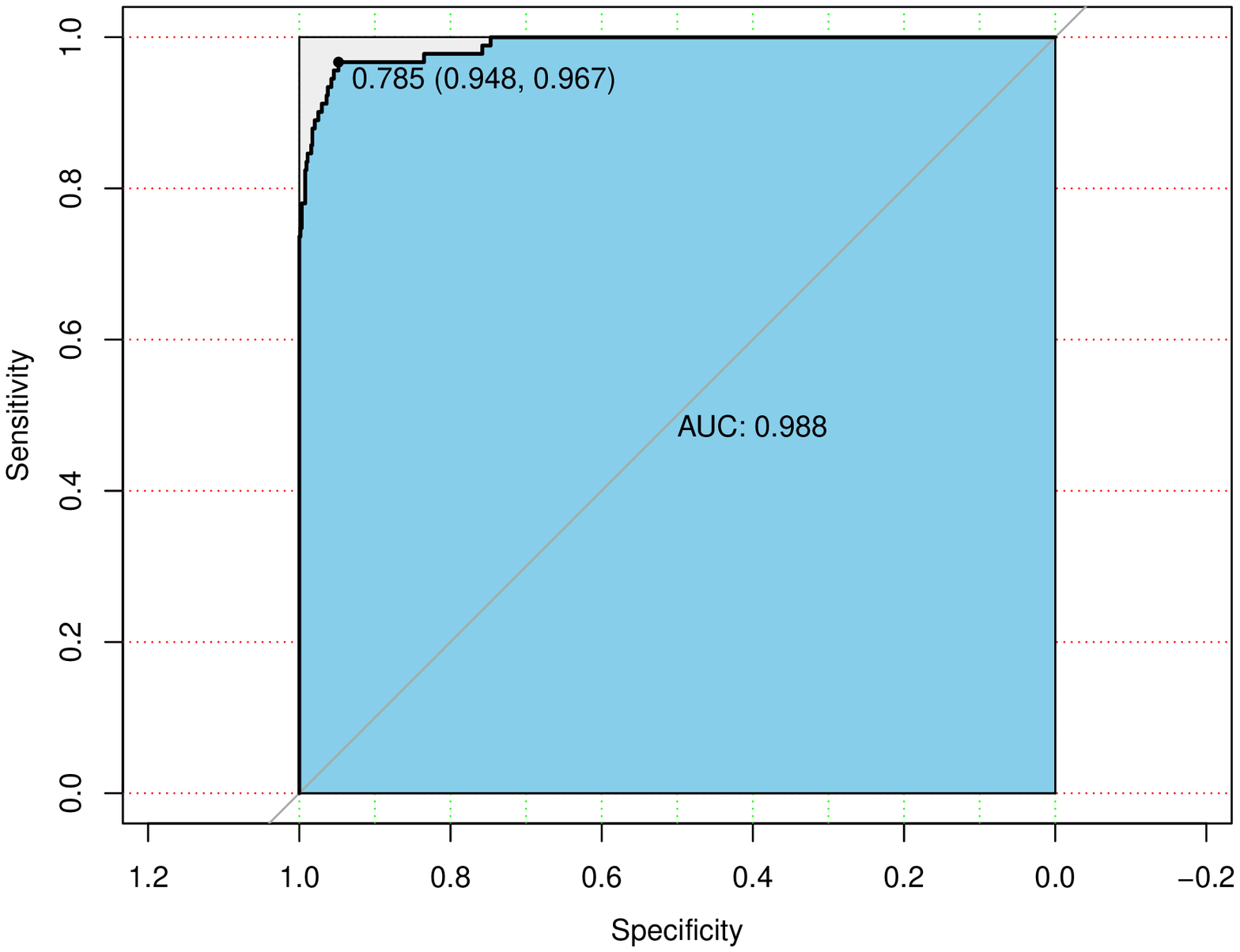}
  \includegraphics[height=5cm,width=7cm]{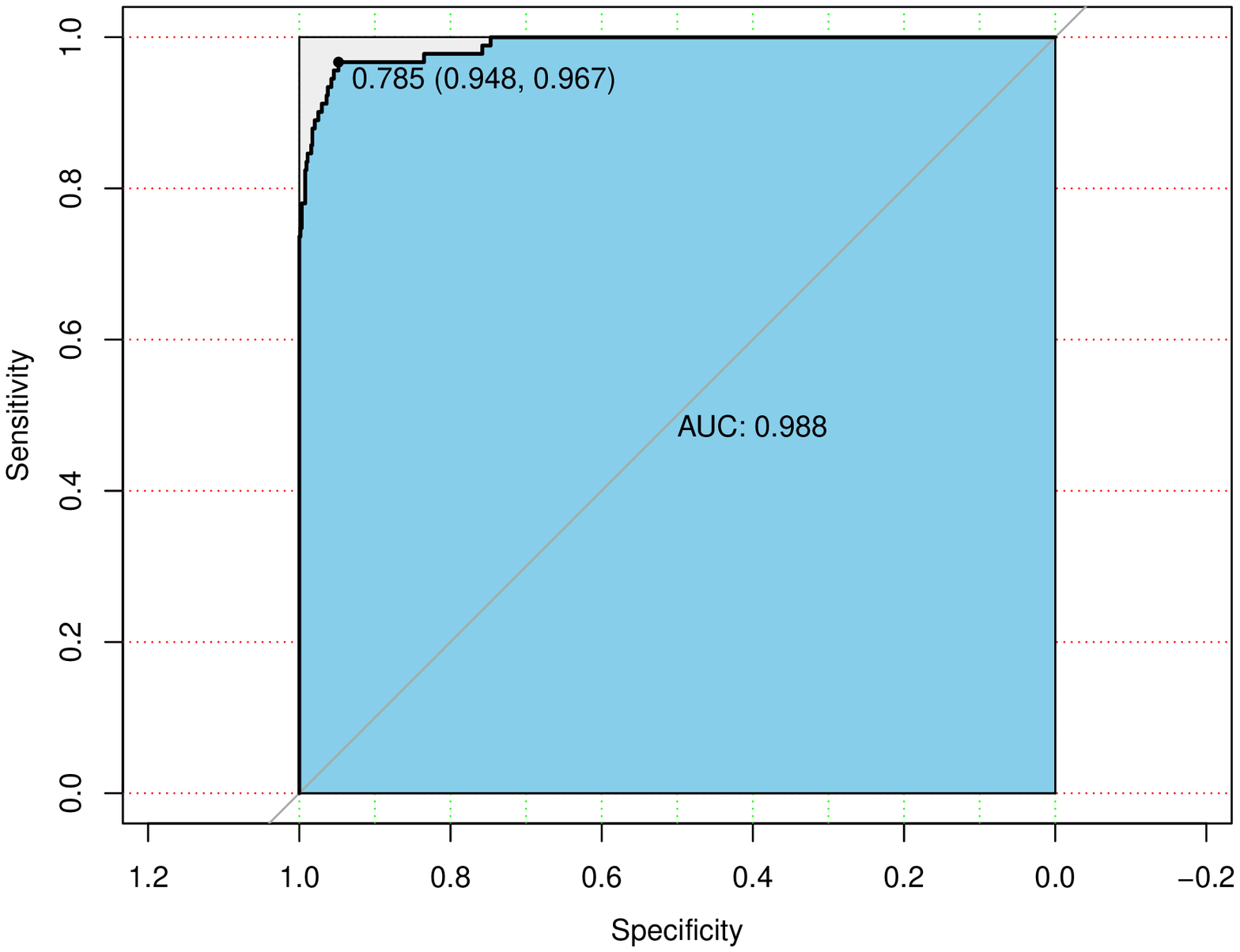}
  \caption{The ROC curves of the ANN classifier with the two best hyper-parameter combination for the validation set in the first step. The text in the figure are the AUC value, the optimal threshold and the corresponding sensitivity and specificity. The different panels correspond to different hyper-parameter combinations}
 \label{fig:fig3}
\end{figure}

\begin{figure}
\centering
\includegraphics[height=5cm,width=7cm]{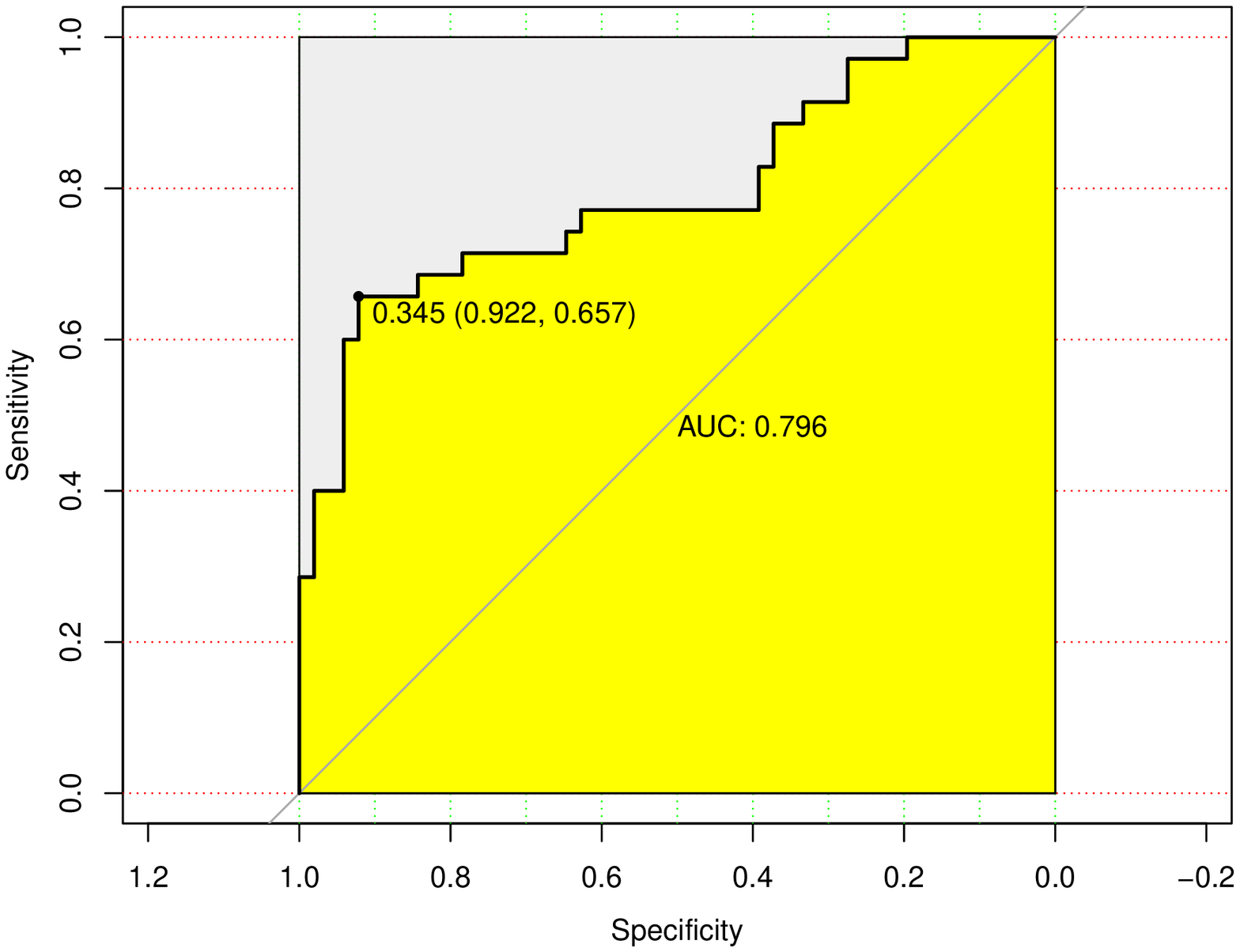}
 \caption{The ROC curves of the ANN classifier with the best hyper-parameter combination for the validation set in the second step. The text in the figure are the AUC value, the optimal threshold and the corresponding sensitivity and specificity.}
 \label{fig:fig4}
\end{figure}

 Compared with RF, ANN is more complicated. The package ``\emph {RSNNS}'' \citep{RSNNS} provides many different types of network structures, including adaptive resonance theory (ART) networks, dynamic learning vector quantization (DLVQ) networks, etc. The most common way to implement ANN classifier is to construct multilayer (MLP) network by calling the function ``\emph {mlp}''. Variable parameters include ``\emph {learnFunc}'', ``\emph {maxit}'' and ``\emph {size}''. The ``\emph {learnFunc}'' represents the used learning function, which contains different network structures, nonlinear activation functions, whether the back propagation is employed and so on. Since the learning function without back propagation is difficult to be stable in a small number of iterations, which may lead to over-fitting, we choose the more common one, ``\emph {BackpropBatch}'', and the parameters of learning function are set to the default value. The ``\emph {maxit}'' represents the maximum of iterations to learn, and the default value is 100. The ``\emph {size}'' is an array that represents the number of hidden layers and the number of neurons per layer. For example, ``\emph c (2,3,4)'' represents three hidden layers, with the number of neurons in each layer being 2, 3, 4, respectively. Considering the limitation of computation, similar to RF, we evaluated the performance of single and double hidden layer classifiers of all the combinations of neurons number per layer in the range of 1 to 15 and ``\emph {maxit}'' in the range of 50 to 150. The hyper-parameter combinations corresponding to the maximum AUC value for the first step are shown in Table \ref{Tab3} and the corresponding ROC curves are shown in Figure \ref{fig:fig3}, while those for the second step are shown in  Table \ref{Tab3} and Figure \ref{fig:fig4}. In the first step, the single hidden layer classifier is better, and there is are two best hyper-parameter combinations, ``\emph {maxit=149}'' and ``\emph {size=9}'', ``\emph {maxit=150}'' and ``\emph {size=9}''. The best AUC is 0.988, and both the thresholds are 0.785 (see  Figure \ref{fig:fig3}). In the second step, the double hidden layer classifier is better, and there is a best hyper-parameter combination, ``\emph {maxit=142}'' and ``\emph {size=c(4,12)}''. The best AUC is 0.796, and the thresholds are 0.345 (see  Figure \ref{fig:fig4}).

 In ANN, we employ a single hidden layer classifier with ``\emph {maxit=149}'' , ``\emph {size=9}'' and threshold of 0.785 in the first step, while a double hidden layer classifier with ``\emph {maxit=142}'' , ``\emph {size=c(4,12)}'' and threshold 0.345 in the second step.

\begin{figure*}
\centering
\includegraphics[height=5cm,width=7cm]{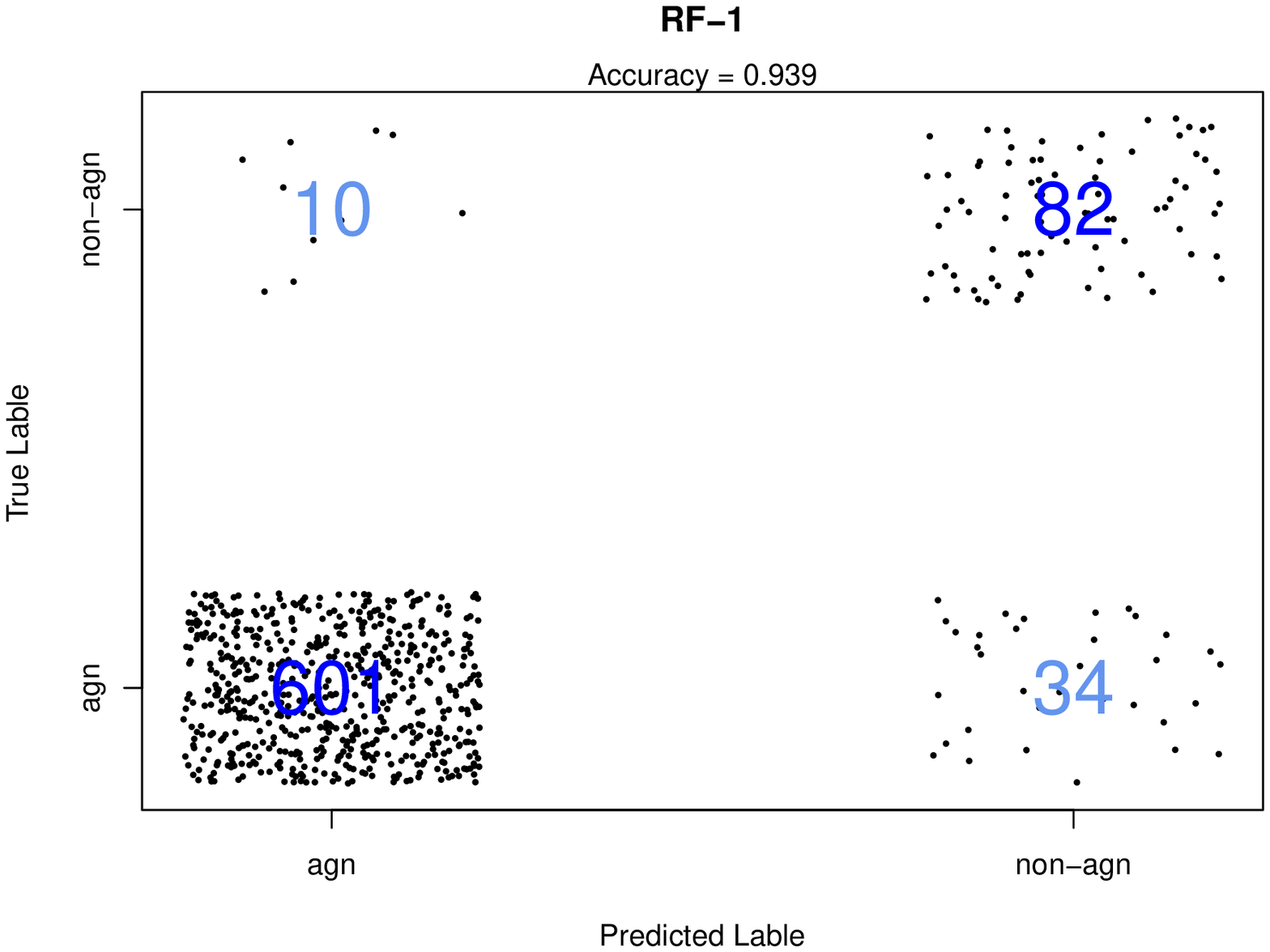}
 \includegraphics[height=5cm,width=7cm]{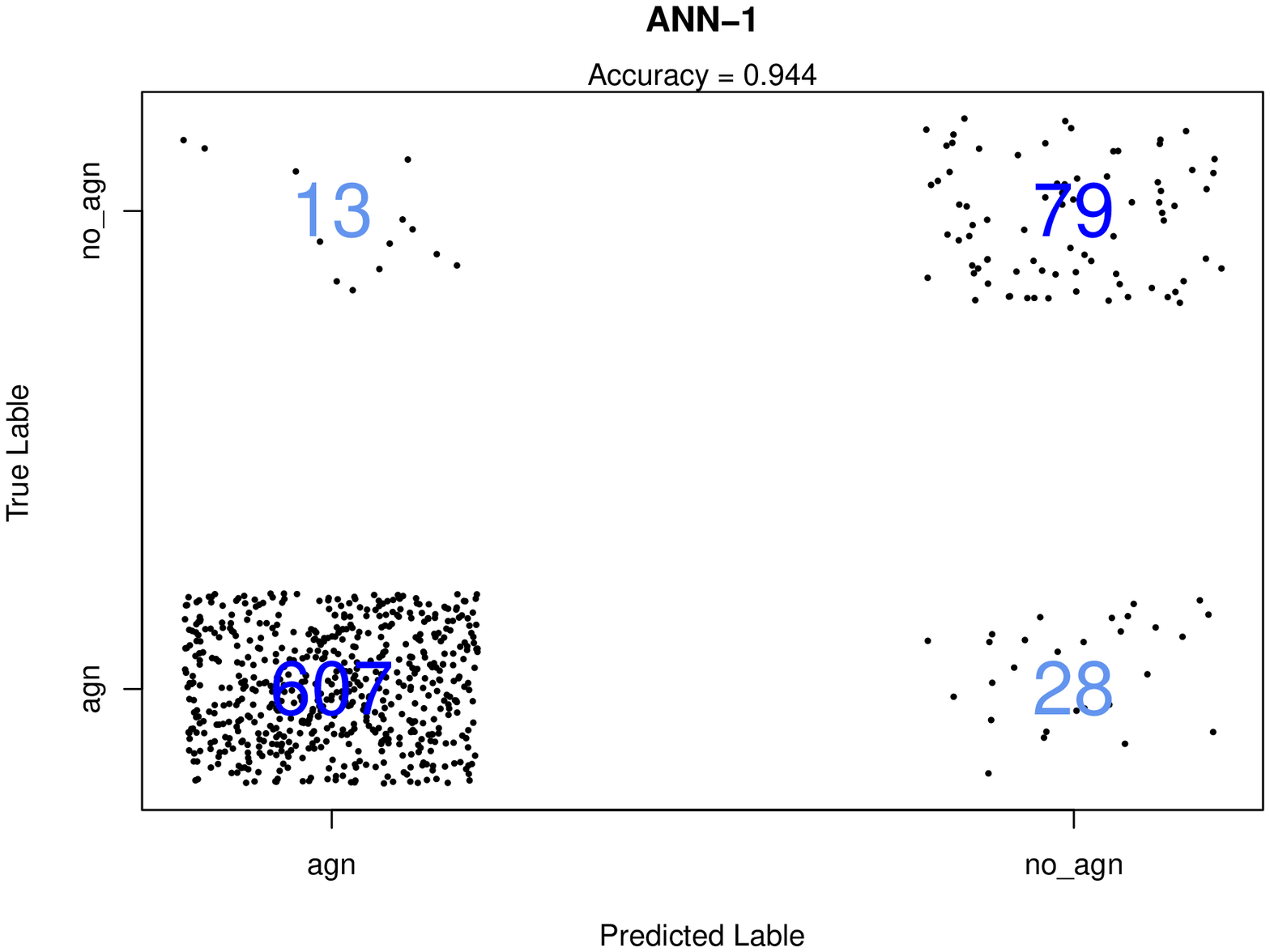}
 \caption{The test confusion matrix of the two classifiers for the first step. The label agn is the positive sample, while the non-agns are the negative samples.}
 \label{fig:fig5}
\end{figure*}

\begin{figure*}
\centering
\includegraphics[height=5cm,width=7cm]{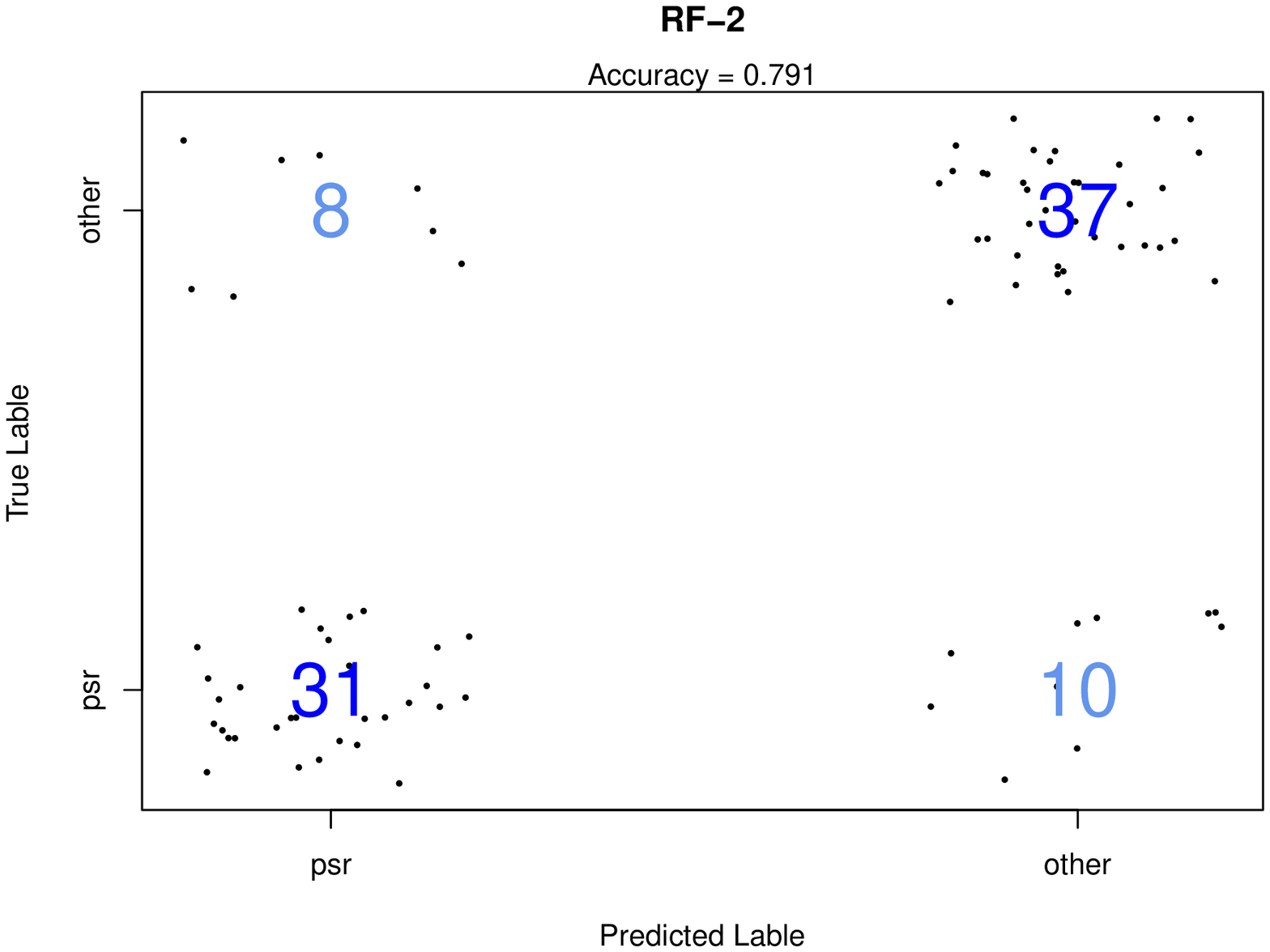}
 \includegraphics[height=5cm,width=7cm]{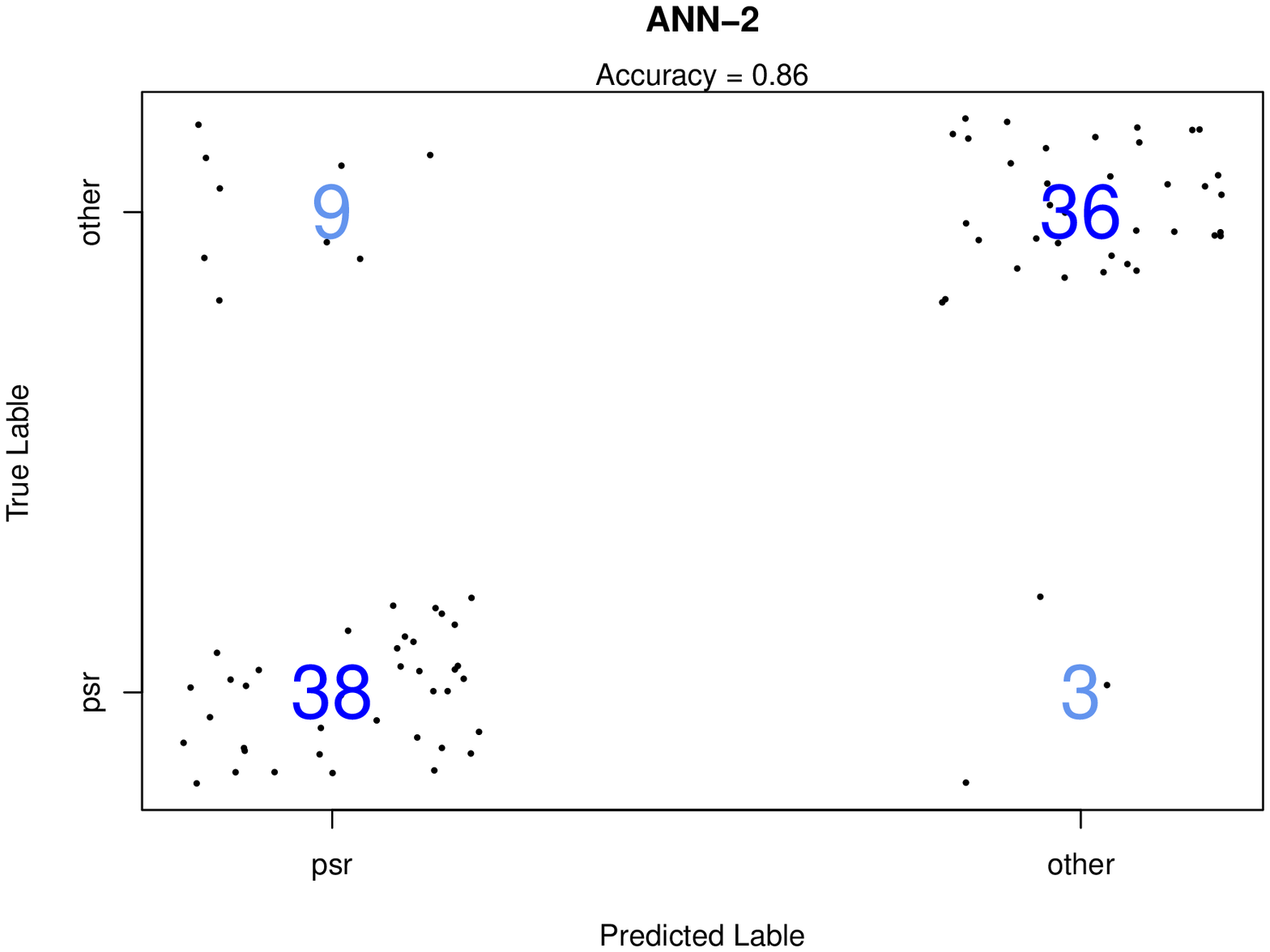}
 \caption{The test confusion matrix of the two classifiers for the second step. The label psr is the positive sample, while the other is the negative sample.}
 \label{fig:fig6}
\end{figure*}

\section{Model testing} \label{sec:test}
\subsection{Individual algorithm results}
Based on the classifier models created (refer to Section 4), we tested their performance with the test set. In the first step, the test set contains 635 AGNs and 92 non-AGNs, and in the second step it includes 41 pulsars and 45 other $\gamma$-ray sources. The test confusion matrixes for the first step are shown in  Figure \ref{fig:fig5}, while the second step is shown in  Figure \ref{fig:fig6}. The performance of the classifiers is shown in Table \ref{Tab4}.

In the first step, the ANN's accuracy was 0.944 slightly higher than the RF's accuracy of 0.939. For the RF algorithm, 34 out of the total of 635 AGNs were misclassified as non-AGNs, while 10 of a total of 92 non-AGNs were misclassified. The sensitivity for non-AGNs was 0.859, and the specificity for the AGNs was 0.956. For the ANN, 28 out of a total of 635 AGNs were misclassified as non-AGNs, while 13 of the total of 92 non-AGNs were misclassified. The sensitivity for the non-AGNs was 0.891, and the specificity for the AGNs was 0.946.

In the second step, the accuracy was not as good. The overall accuracies of two algorithms were 0.791, 0.860, respectively. The RF algorithm has a high sensitivity of 0.822 for the other gamma-ray sources and less misclassification (8 out of 45). The specificity for pulsars was 0.791 and 10 of a total of 41 pulsars were misclassified as other sources.In contrast, the ANN has high specificity up to 0.927 of the pulsars and less misclassification (3 out of 41). The sensitivity for the other category was 0.800, and 9 of a total of 45 other sources were misclassified as pulsars.

\subsection{Composition algorithm results}
When combining the two algorithms, we are guided by the principle of common identification, that is, we obtained the classification results only when unassociated sources are classified as the same by both classifiers. When the source classification results of the two classifiers are inconsistent, we consider the sources to be the uncertain type (label as ``unc''). For example, the source numbered as 4FGL J0531.7+1241c is obtained as uncertain type, while it is evaluated as an AGN in ANN classifier and evaluated as an other $\gamma$-ray source in RF classifier. Hence, the accuracy is improved, although the number of candidates is reduced \citep[e.g.,][]{2019ApJ...887..134K}. The combined test results of the two algorithms are shown in Table \ref{Tab5}. For the AGNs, there are only nine misclassifications of the 614 candidates obtained, and the overall accuracy is over $98\%$ . In the case of pulsars and other sources, the overall accuracies were 0.886 and 0.914, respectively. There are also some sources of indeterminate type.

\begin{figure*}
\centering
 \includegraphics[height=5cm,width=7cm]{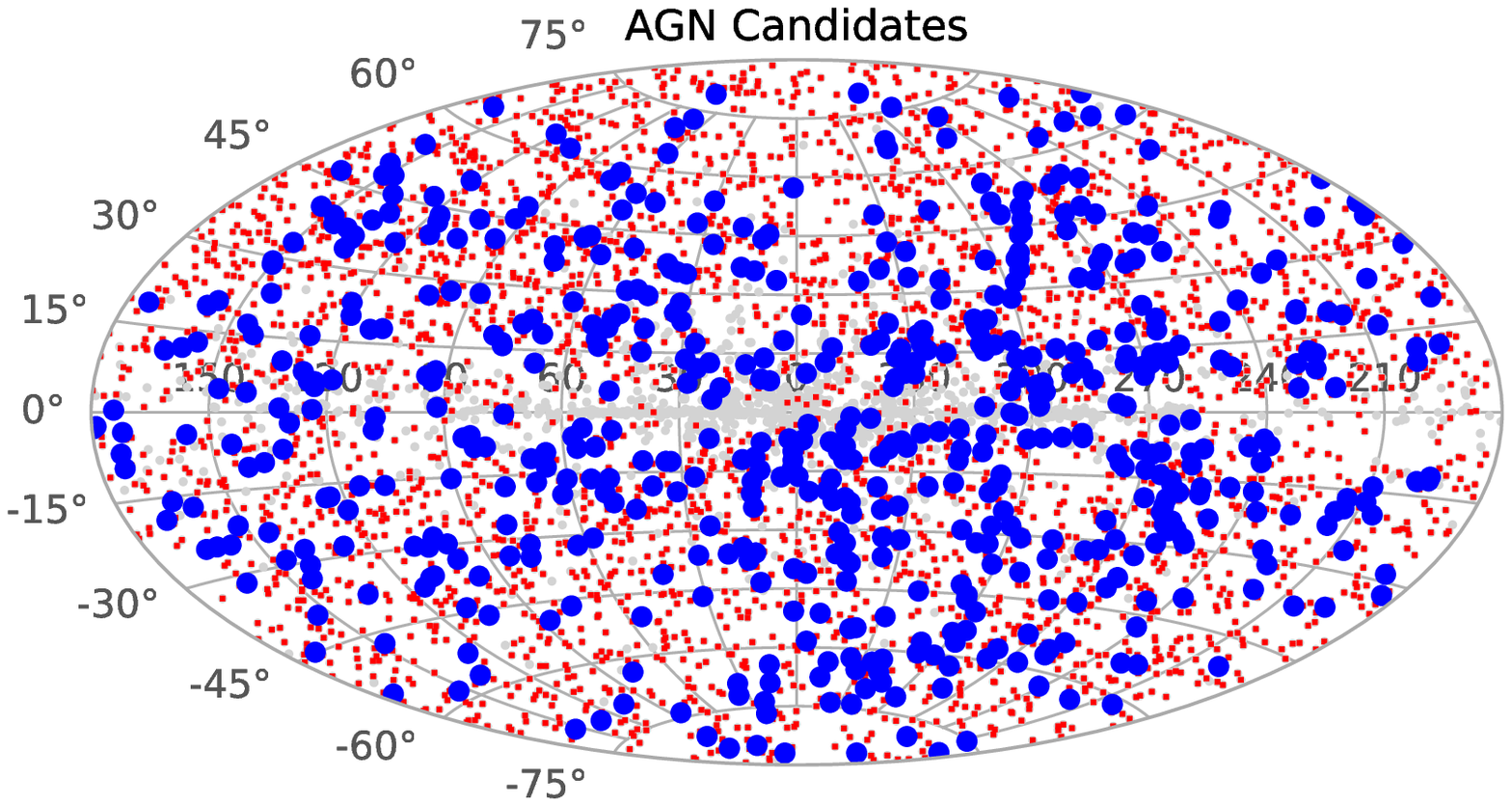}
   \includegraphics[height=5cm,width=7cm]{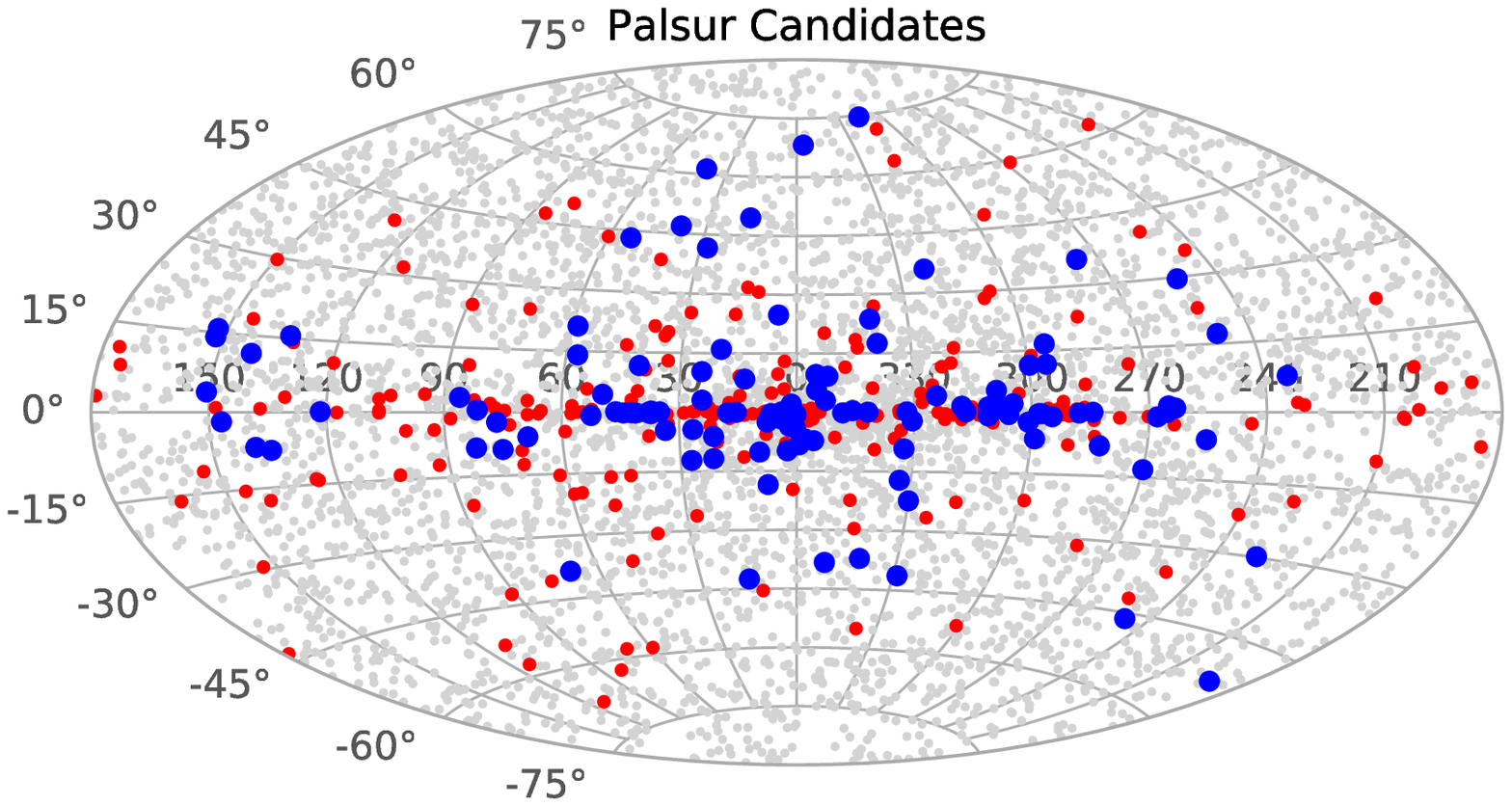}
 \caption{All sky distribution of AGN (left) and pulsar (right) candidates in Galactic coordinate. The blue symbol represents candidates we obtained, the red symbol represents the sources of AGN or pulsar identified or associated in 4FGL and gray symbol represents all $\gamma$-ray sources in 4FGL. }
 \label{fig:fig7}
\end{figure*}

Then, we employ the classification model to the 4FGL catalog's dataset of 1336 unassociated sources. The ANN classifier gives 911 AGNs, 166 pulsars, and 259 other gamma-ray candidates. The RF classifier gives 585 AGNs, 175 pulsars, and 576 other gamma-ray candidates. Combining the results of the two classification algorithms, we obtain 583 AGN candidates, 115 pulsar candidates, and 154 other gamma-ray candidates (see Table \ref{Tab6}). Figure \ref{fig:fig7} shows the distribution of the AGN and pulsar candidates over the sky. We find that most of pulsar candidates are located near the galactic plane. 74 pulsar candidates are located at GLAT $\left|{b}\right| \leq 10^{\circ}$, 11 candidates are located at GLAT $10^{\circ} \leq \left|{b}\right| \leq 15^{\circ}$. The distribution is consistent with the identified pulsars. However, the AGN candidates are dispersed the all sky. Just only 108 AGN candidates are located at GLAT $\left|{b}\right| \le 10^{\circ}$. Since the high density distribution of the sources and the bright background near the galactic plane, it is considered that the AGN candidates of low GLAT are difficult to identify.

\begin{table*}
\centering
\caption{Test results for two classifiers}\label{Tab4}
\begin{tabular}{ccccccccc} 
\hline\hline
&&\multicolumn{3}{c}{First step}   &&\multicolumn{3}{c}{Second step}   \\
\cline{3-5} \cline{7-9}			
Classifier      &&Sensitivity&Specificity&Accuracy    &&Sensitivity&Specificity&Accuracy\\
\hline
RF	&&0.891	&0.946	&0.939	&&0.822	&0.756	&0.791\\
ANN	&&0.859	&0.956	&0.944	&&0.800	&0.927	&0.860\\
\hline
\end{tabular}\\
{Note: Column 1: Classification methods used in this paper.
Columns 2 - 4 show the test results for the first step: the sensitivity for the non-AGNs and the specificity for AGNs, and overall accuracy, respectively.
Columns 5 - 7 reports the test results for the second step: sensitivity for other $\gamma$-ray sources,  specificity for pulsars and overall accuracy, respectively.
}
\end{table*}

\begin{table*}
\centering
\caption{Test results for classifiers combined}\label{Tab5}
\begin{tabular}{ccccccccc} 
\hline\hline
Class &&Label &&Count	&&Errors &&Overall accuracy\\
\hline
AGN&&agn&&605&&9&&0.985\\
Pulsar&&psr&&35&&4&&0.886\\
Other $\gamma$-ray source&&other&&35&&3&&0.914\\
\hline
\end{tabular}\\
{Note: Columns 1 and 2: Source classes and labels.
Columns 3 and 4: Source count and the number of misclassifications for each label when combined two classifiers.
Column 5: Overall accuracy for each label when combining the two classifiers.
}
\end{table*}

\begin{table}
\centering
\caption{The classification of unassociated sources\label{Tab6}}
\resizebox{\textwidth}{!}{
\begin{tabular}{ccccccccccccc}
\hline
4FGL$\_$name & R.A. & Dec&
$P_{RF1}$   & $P_{RF2}$& $C_{RF}$ &
$P_{ANN1}$& $P_{ANN2}$ & $C_{ANN}$&
$C_{com}$ & $A_{name}$ & $C_{LR-P}$
& $C_{RF-P}$\\
\hline
$\cdots$&$\cdots$&$\cdots$&$\cdots$&$\cdots$&$\cdots$&$\cdots$&$\cdots$&$\cdots$&$\cdots$       \\
4FGL J0530.0-6900e	&	82.50	&	-69.00	&	0.09	&	0.02	&	other	&	0.59	&	0.10	&	other	&	other	&	3FGL J0524.5-6937 	&	AGN	 &	AGN	\\
4FGL J0531.7+1241c	&	82.94	&	12.69	&	0.63	&	0.32	&	other	&	0.81	&		&	agn	&	unc	&	                  	&		&		\\
4FGL J0531.8-6639e	&	82.97	&	-66.65	&	0.83	&		&	agn	&	0.98	&		&	agn	&	agn	&	3FGL J0525.2-6614 	&	AGN	&	AGN	\\
4FGL J0532.6+3358 	&	83.17	&	33.98	&	0.82	&		&	agn	&	0.99	&		&	agn	&	agn	&	                  	&		&		\\
4FGL J0533.6+5945 	&	83.42	&	59.76	&	0.28	&	0.98	&	psr	&	0.41	&	0.94	&	psr	&	psr	&	3FGL J0533.2+5944 	&	PSR	&	AGN	\\
4FGL J0533.9+2838 	&	83.48	&	28.64	&	0.83	&		&	agn	&	1.00	&		&	agn	&	agn	&	                  	&		&		\\
4FGL J0534.0+3746c	&	83.51	&	37.77	&	0.71	&	0.39	&	other	&	1.00	&		&	agn	&	unc	&	                  	&		&		\\
4FGL J0534.2+2751 	&	83.57	&	27.86	&	0.99	&		&	agn	&	0.97	&		&	agn	&	agn	&	                  	&		&		\\
4FGL J0535.1-5422 	&	83.78	&	-54.38	&	0.93	&		&	agn	&	0.97	&		&	agn	&	agn	&	                  	&		&		\\
4FGL J0535.3+0934 	&	83.84	&	9.58	&	0.98	&		&	agn	&	1.00	&		&	agn	&	agn	&	                  	&		&		\\
4FGL J0536.1-1205 	&	84.03	&	-12.09	&	0.90	&		&	agn	&	0.92	&		&	agn	&	agn	&	                  	&		&		\\
4FGL J0537.5+0959 	&	84.38	&	9.99	&	0.98	&		&	agn	&	1.00	&		&	agn	&	agn	&	3FGL J0537.0+0957 	&	AGN	&	AGN	\\
4FGL J0538.9+3549c	&	84.74	&	35.83	&	0.33	&	0.09	&	other	&	0.53	&	0.20	&	other	&	other	&	                  	&		 &		\\
4FGL J0539.2-6333 	&	84.82	&	-63.55	&	1.00	&		&	agn	&	1.00	&		&	agn	&	agn	&	                  	&		&		\\
4FGL J0540.0-7552 	&	85.01	&	-75.88	&	0.90	&		&	agn	&	1.00	&		&	agn	&	agn	&	3FGL J0539.9-7553 	&	AGN	&	AGN	\\
4FGL J0540.2+0655 	&	85.05	&	6.92	&	1.00	&		&	agn	&	0.99	&		&	agn	&	agn	&	                  	&		&		\\
4FGL J0540.6+5540 	&	85.17	&	55.67	&	1.00	&		&	agn	&	1.00	&		&	agn	&	agn	&	                  	&		&		\\
4FGL J0540.7+3611 	&	85.18	&	36.20	&	0.53	&	0.41	&	other	&	0.40	&	0.66	&	psr	&	unc	&	                  	&		&		 \\
4FGL J0543.5-8741 	&	85.89	&	-87.69	&	1.00	&		&	agn	&	1.00	&		&	agn	&	agn	&	3FGL J0542.2-8737 	&	AGN	&	AGN	\\
4FGL J0543.9-0418 	&	85.98	&	-4.31	&	0.81	&		&	agn	&	0.98	&		&	agn	&	agn	&	                  	&		&		\\
4FGL J0544.4+2238 	&	86.11	&	22.64	&	0.71	&	0.21	&	other	&	0.98	&		&	agn	&	unc	&	3FGL J0544.7+2239 	&	AGN	&	AGN	\\
4FGL J0544.8+5209 	&	86.22	&	52.16	&	1.00	&		&	agn	&	1.00	&		&	agn	&	agn	&	                  	&		&		\\
4FGL J0545.7+6016 	&	86.44	&	60.27	&	0.16	&	0.93	&	psr	&	0.58	&	0.94	&	psr	&	psr	&	3FGL J0545.6+6019 	&	PSR	&	PSR	\\

$\cdots$&$\cdots$&$\cdots$&$\cdots$&$\cdots$&$\cdots$&$\cdots$&$\cdots$&$\cdots$&$\cdots$       \\
\hline
\end{tabular}}\\
{Note:Column 1 shows the 4FGL names. The right ascension and declination of sources are listed in Columns 2-3, respectively.
Columns 4-5 report the score given by ANN classifier for the first ($P_{ANN1}$) and second ($P_{ANN2}$) step. Sources with a step 1 score below the threshold 0.789 are classified as non-AGNs and brought into the step 2 classification. The classification ($C_{ANN}$) given in the ANN is listed in Column 6.
Columns 7-8 report the scores given by the RF classifier for the first ($P_{RF1}$) and second  ($P_{RF2}$) step. Sources with a step 1 score below the threshold 0.739 are classified as non-AGNs and brought into the step 2 classification. The classification ($C_{RF}$) given in the RF is listed in Column 9.
Column 10 shows the classification results of the two algorithms  combined (``unc'' means uncertain source).
Column 11 shows the associated name ($A_{name}$) in the other FGL. The cross-matching results for the 3FGL catalog's unassociated sources classification results \cite{2016ApJ...820....8S} obtained using logistic regression ($C_{LR-P}$) and random forest ($C_{RF-P}$) are listed in Column 12 and 13, respectively. Table \ref{Tab6} is published in its entirety in the machine-readable format. A portion is shown here for guidance regarding its form and content. (This table is available in its entirety in machine-readable form. e.g., Table6\_result.xlsx)
}
\end{table}

\section{Conclusion and Discussion} \label{sec:Conclusion and Discussion}
In this work, we attempt to search for AGN and pulsar candidates in the 4FGL catalog's unassociated samples based on two supervised learning methods. We do not focus on the specific physical mechanism. To accommodate the unbalanced sample, we divide the classification process into two steps. Firstly, we use the RF and ANN methods with 20 features selecting by the K-S test to select AGN candidates in all of the unassociated samples. Then, we utilize the same methods with eight different features to select pulsar candidates in the remaining non-AGN samples for the second step. By finding the optimal combination of hyper-parameters to optimize the algorithm, we test the performance of our model (accuracy, sensitivity, etc.), and evaluate the labels of the 1336 unassociated samples. The accuracy obtained in the first step is about $95\%$, and in the second step, the obtained overall accuracy is approximately $80\%$. Finally, we obtain 583 AGN candidates, 115 pulsar candidates, 154 other type of candidates, and 484 of uncertain type by combining the results of the two classifiers.

The current context provides the coordinates and the all-sky map of the obtained AGN and pulsar candidates. Meanwhile, the probabilities given by different classifiers of each source are also shown (see Table \ref{Tab5}). These could help us to the sky survey, as well as to the further study on Fermi unassociated sources by the investigators. We note that AGNs and non-AGNs differ in spectral shape, variability, overall integral flux and flux of partial band (such as, from 300 MeV to 10 GeV), which is related to the high-energy emission mechanism of AGNs (e.g., \citealt{2016A&A...585A...8Z,2016MNRAS.457.3535Z,2017ApJS..228....1Z}). On the other hand, the pulsar and non-pulsar are quiet different in spectral shape and higher energy band flux (such as, from 30 to 300 GeV), that result from the unique high-energy emission mechanism of $\gamma$-ray pulsars (e.g., \citealt{1986ApJ...300..500C,1996ApJ...470..469R,2014Sci...344..159R}).

\begin{figure*}
\centering
 \includegraphics[height=5cm,width=7cm]{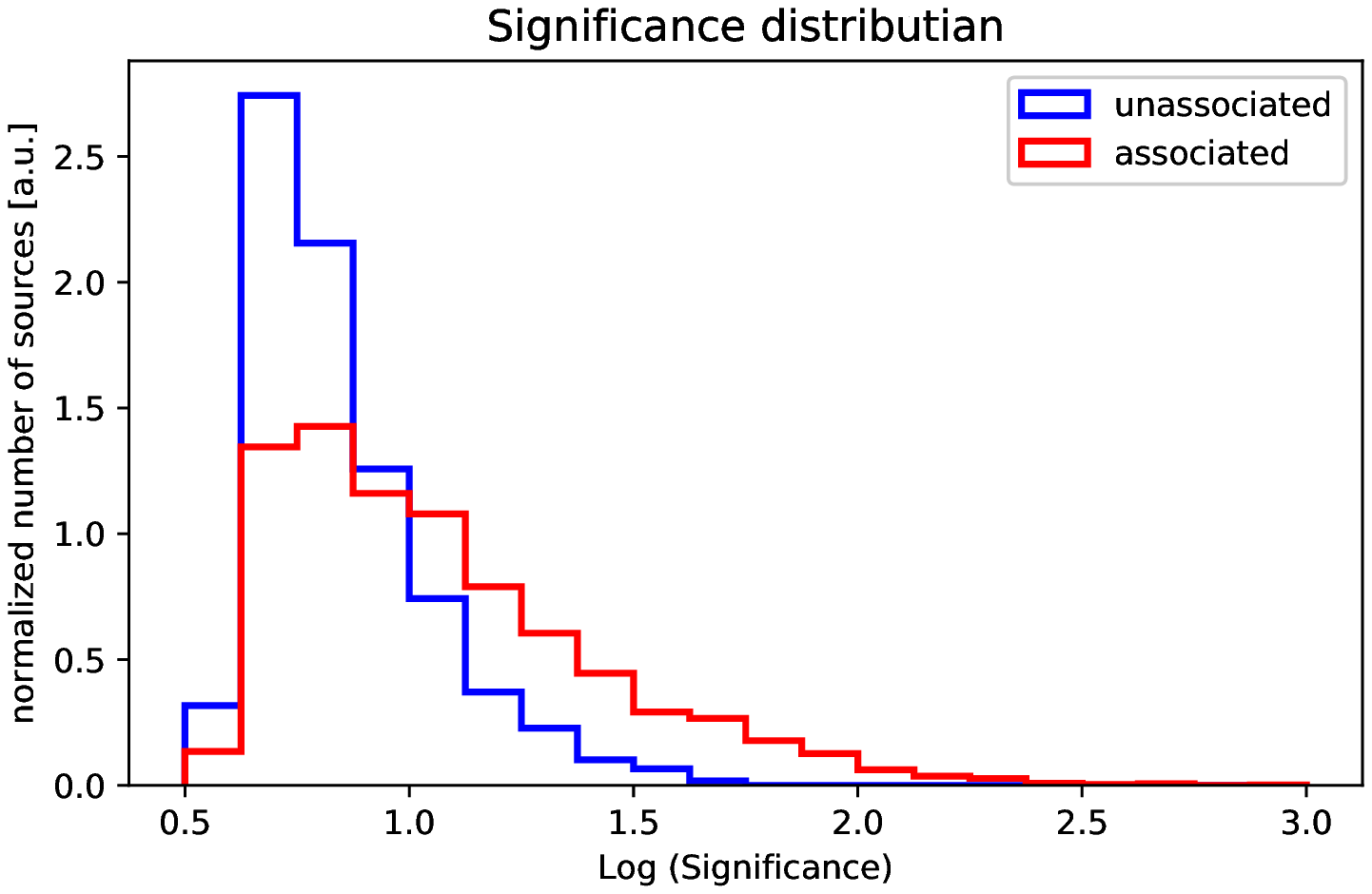}
   \includegraphics[height=5cm,width=7cm]{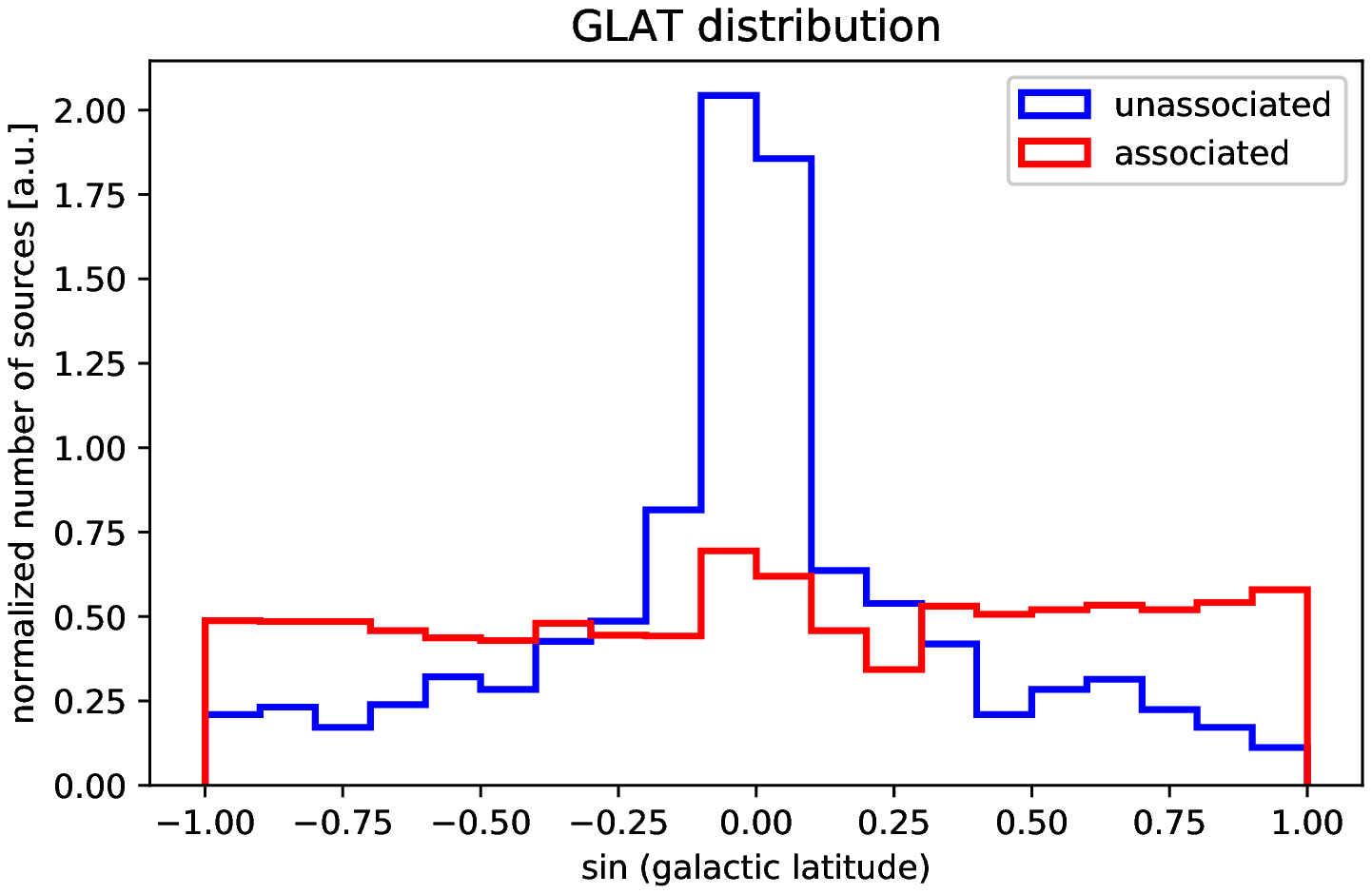}
 \caption{The normalized source distribution in significance (left) and Galactic latitude (right).}
 \label{fig:fig8}
\end{figure*}

The basis for SML for classification is the training samples and the predicted sample to be classified accord with the same distribution in the multi-dimensional feature space. When the distributions were different, we encounter the potential problem that the classifier does not perform as well on the unassociated samples as it does on the test samples, which also known as covariate shift or sample selection bias in astronomy (See \citealt{2012ApJ...744..192R,2012adm..book..213R,2020MNRAS.tmp..163L} for the more detail discussions). In the classification of  Fermi unassociated sources, the covariate shift exists when comparing the distribution differences of some features between 3FGL and 4FGL(e.g., $Variability\_Index$, see \citealt{2020MNRAS.tmp..163L}).
As observations advance, the features are changing with longer exposure, improvement of observational, statistical methods and the identification or associate of partial sources.
Just as the brighter training samples in variable star classification lead to sample selection bias in the classification of fainter stars \citep{2012ApJ...744..192R,2012adm..book..213R}, there are systematic differences between fermi associated and unassociated sources. Bright $\gamma$-ray sources are more likley to be bright at other wavelengths
(radio, optical, X-ray) and therefore more likely to be detected in multiwavelength catalogs that are used to associate $\gamma$-ray sources. The sources of the associated sample that are used for training and testing the performance of the classification algorithms are generally brighter and detected at higher significance level. On the contrary, the unassociated source were non-significant. This may lead to the potential problem that the classifier model is not ideal for predicted target, even though it performs well on the test samples. Similarly, the systematic differences are reflected in the coordinate space. The unassociated samples were biased towards to the sources near the galactic plane, while the associated were widely distributed throughout the all-sky, especially in region with high Galactic latitude. The large number of sources and highlighted backgrounds on the Galactic disk increase the difficulty of source detection.The source distribution in the significance space and Galactic latitude are shown in Figure \ref{fig:fig8}.

\begin{figure*}
\centering
 \includegraphics[height=5cm,width=7cm]{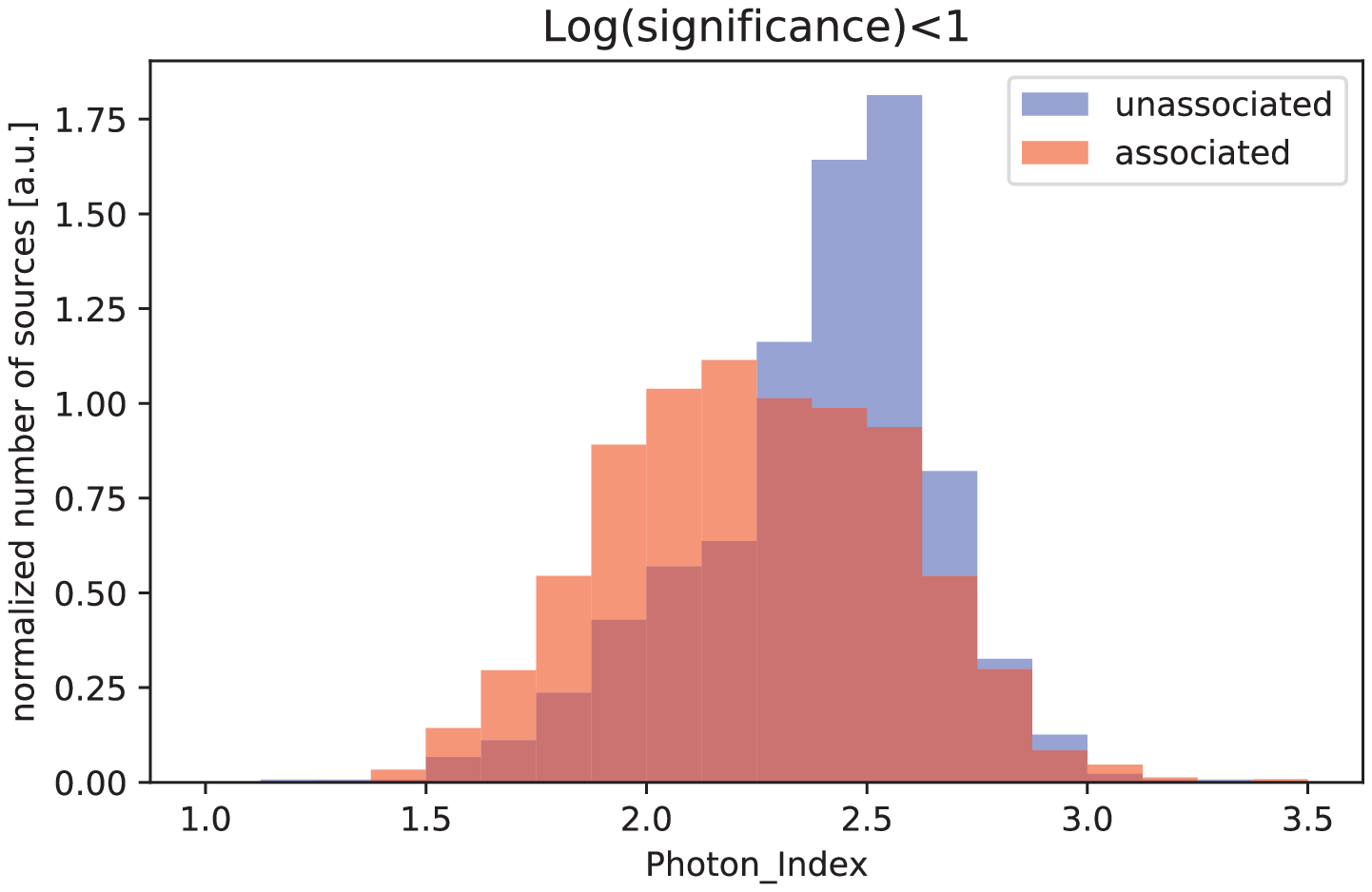}
   \includegraphics[height=5cm,width=7cm]{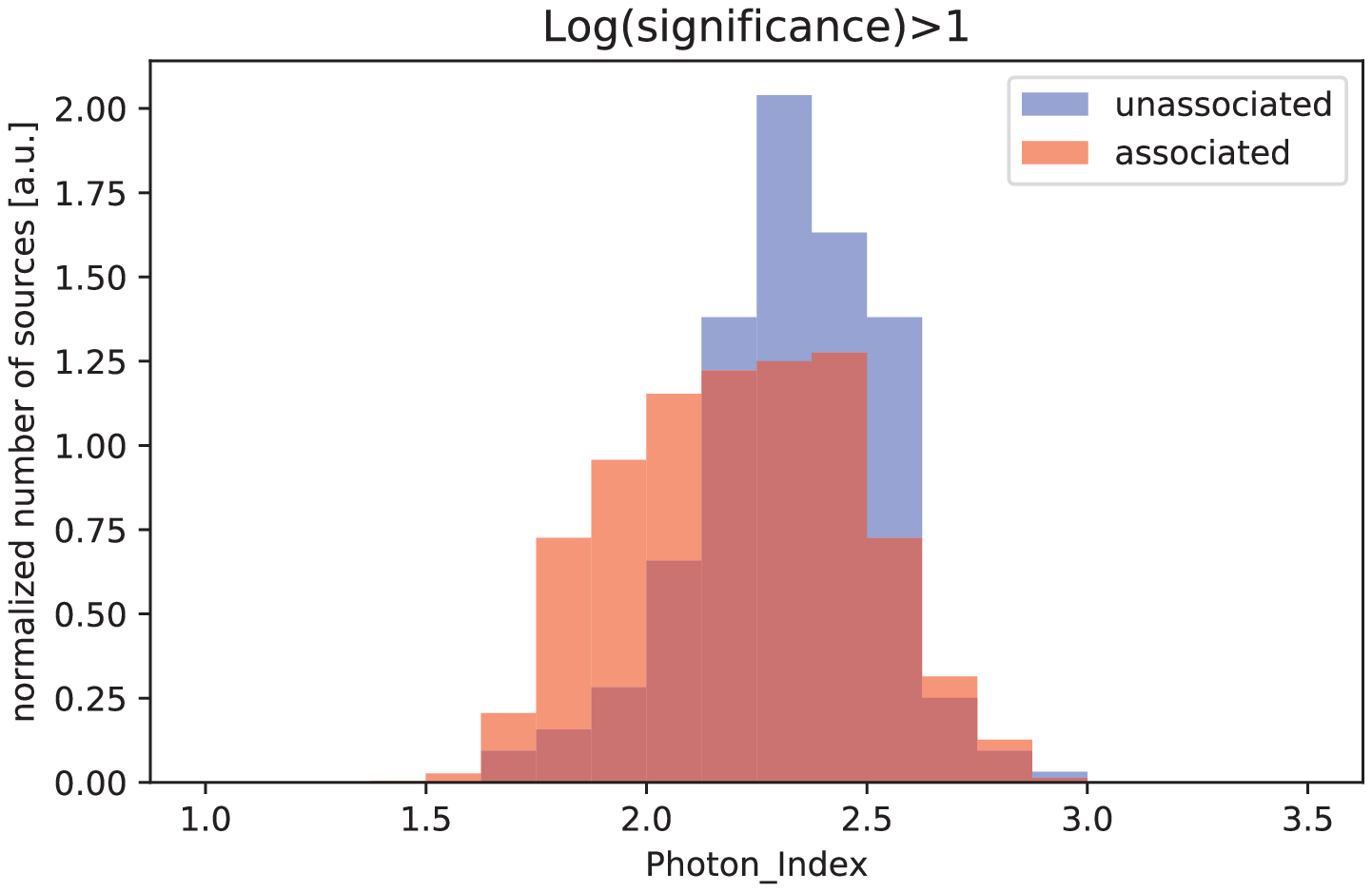}
     \includegraphics[height=5cm,width=7cm]{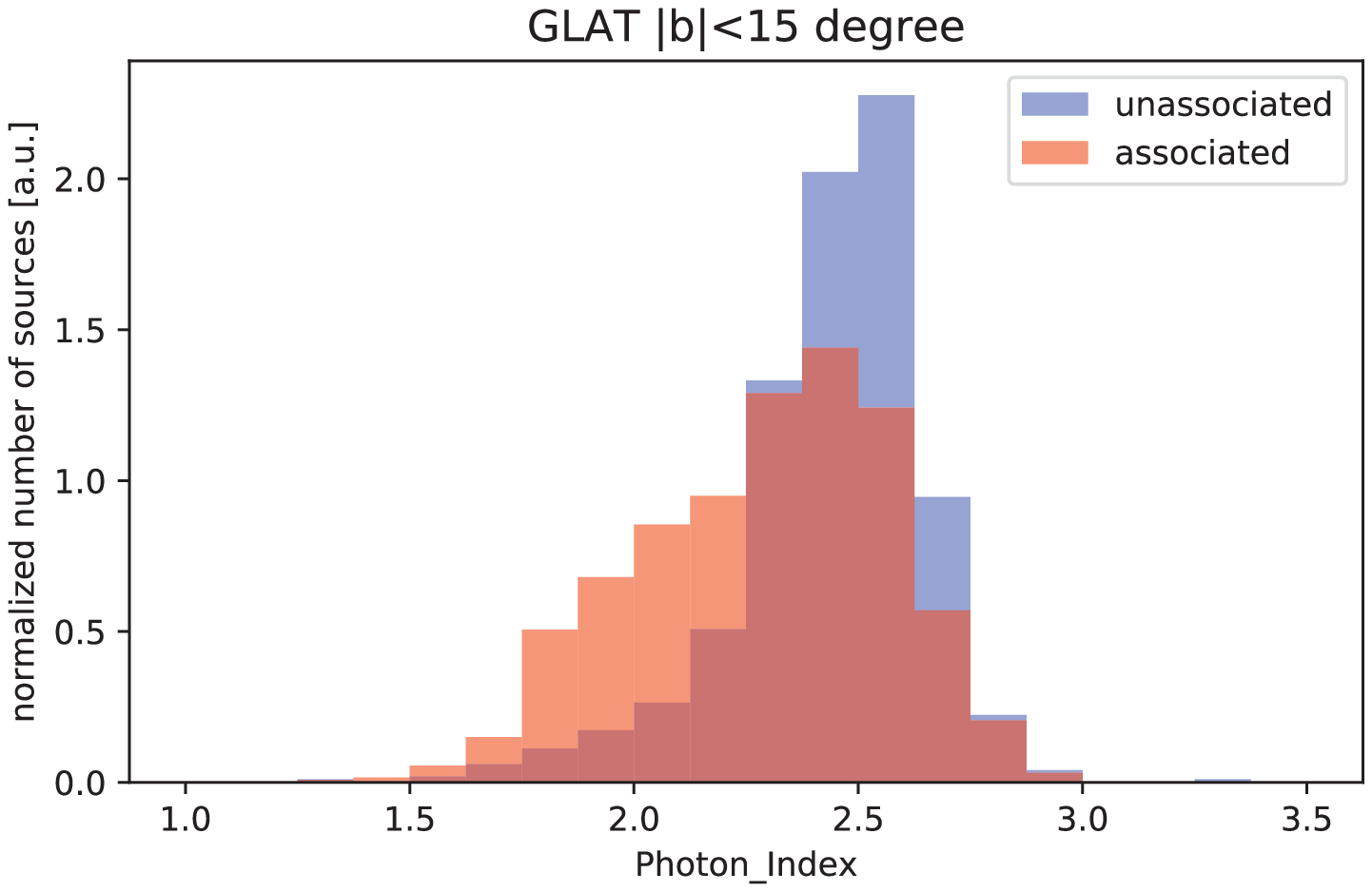}
       \includegraphics[height=5cm,width=7cm]{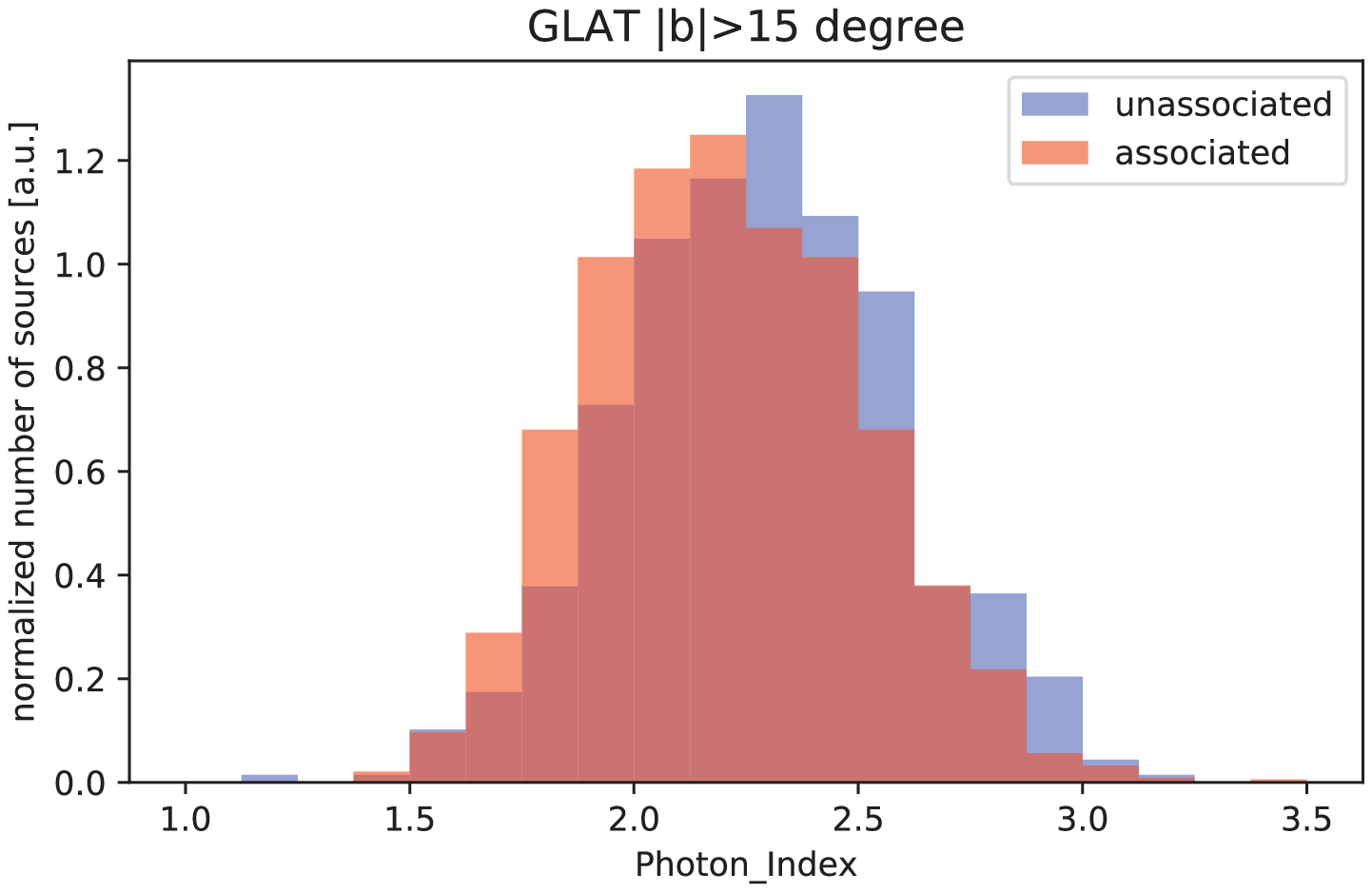}

 \caption{The normalized distribution of photon spectrum index of darker sources, brighter sources, lower Galactic latitude sources and higher Galactic latitude sources. The darker part is the coincidence region of associated and unassociated samples.}
 \label{fig:fig9}
\end{figure*}

For the purpose of clarifying the influence of systematic differences in distribution on the classification, we divided the Fermi sources into four groups: brighter source, darker source, higher Galactic latitude sources and lower Galactic latitude sources.Taking $Photon\_index$ as an example, the normalized distribution diagram is given in Figure \ref{fig:fig9}. The spectral indexes of associated and unassociated sources located at higher Galactic latitude are similar in comparison with those of low Galactic latitude and therefore it is a ``cleaner'' dataset for using feature $Photon\_index$ to classification. For significance, the distribution differences of associated and unassociated samples in both brighter  and darker source are large, while the difference proportion in the brighter source is slightly smaller.

Due to the limitation of astronomical observation, sample selection bias is almost inevitable. A simple classifier with few features reduces the possibility of covariant shift \citep{2020MNRAS.tmp..163L}. Using hardness ratios instead of direct observation like individual fluxes and variability index to keep information about the spectral shape, or modifying the  observations to obtain more intrinsic features might solve the problem. Grouping the prediction samples and searching for suitable training samples to refine the classification process is less likely to encounter the problem of sample selection bias.

In recently, \cite{2016ApJ...820....8S} divided all of the 3FGL catalog's sources into AGNs and pulsars based on the LR and RF algorithms. Cross-matching the 4FGL predictions (1336 unassociated sources) from the present work and 3FGL predictions (3033 sources, see \citealt{2016ApJ...820....8S}\footnote{Also see \url{https://www.physics.hku.hk/~pablo/pulsarness/Step_08_Results.html}}), we obtained 334 common sources (see Table \ref{Tab6}). In the 3FGL predictions of common sources, 146 sources were classified as AGNs and 188 sources were classified as pulsars in LR$^{P,}$\footnote{The Logistic regression and random forest model used in \citealt{2016ApJ...820....8S}}; 163 sources were classified as AGNs and 171 sources were classified as pulsars in their ${\rm RF}^P$ based classifier. In our predictions of classifiers combination, 96 sources belonged to the AGN type, 73 belonged to the pulsar type and the number of sources classified as other or uncertain type are 24 and 141, respectively. Cross-matching the results (see Table \ref{Tab7}), the majority of sources (approximately $89\%$) had the same classification for sources of AGN and pulsar types in our predictions.
In \cite{2020MNRAS.tmp..163L}, they searched 20 millisecond pulsar candidates from the 4FGL unidentified sources using a two-layer cascade method prompted by investigating the factors affecting machine learning classifications. Cross-matching the 20 millisecond pulsar candidates given by \cite{2020MNRAS.tmp..163L}, 9 sources are evaluated as pulsars, 2 sources are classified as AGNs, and 9 sources are uncertain type in our predictions. In addition, we note that the 9 pulsar candidates have higher significance in their results, while 2 AGNs with lower significance.
Most of our predictions are consistent with other previous results. However, the predictions of a fraction of sources are inconsistent, and the evaluation of their true classification needs further study in the future.

\begin{table*}{}
\centering
\caption{Comparison of 3FGL and 4FGL results}\label{Tab7}
\begin{tabular}{cccccccc}
\hline\hline
~&~&~&~&\multicolumn{4}{c}{{Obtained predictions}} \\
\cline{5-8}
method&Label &Count$^b$ &&agn&psr&other&unc\\
\hline
&Count$^a$&334&&96&73&24&141\\
\hline
LR$^P$&agn&146&&84&6&4&52\\
&psr&188&&12&67&20&89\\
\hline
RF$^P$&agn&163&&87&9&4&63\\
&psr&171&&9&64&20&78\\
\hline
\end{tabular}\\
{Note: Column 1: The classifiers obtained in \citealt{2016ApJ...820....8S}.
Columns 2 and 3: The classification results for 356 common sources from \citealt{2016ApJ...820....8S} , where the row count$^a$ shows our classification results for 356 common sources.
Columns 4-7: Cross comparison results.
}
\end{table*}

We have tried to put all of the unassociated samples (i.e., 1336) into the algorithms at the same time and classify them into three types directly. Although an overall accuracy of over 0.9 can be obtained (see Table \ref{Tab8}), the approach has several limitations. Firstly, the result is unstable, especially for the other type, and the results given by various classifiers are quite different. Secondly, the imbalance of the samples reduces the accuracy. Almost no predicting sample is classified as the other type, mainly resulting from fewer other type samples with insignificant characteristics. The presence of more AGN type of training samples leads to more unassociated samples to be evaluated as AGN types. For unbalanced samples, this can result in higher accuracy but it doesn't mean the classifier a good classifier. Thirdly, there is a large difference in the selection of suitable features for different samples. For example, based on the results of the K-S test, ``logF4'' is the best feature in evaluation of the AGNs and non-AGNs, but it is not a good feature in the discrimination of pulsars and non-pulsars. In order to obtain a higher confidence level, we employ a step-by-step feature selection and classification approach at the expense of higher computational demands.

\begin{table*}
\centering
\caption{Results from two supervised classifiers for simultaneous classification of three different types}\label{Tab8}
\resizebox{\textwidth}{!}{
\begin{tabular}{cccccccccccc} 
\hline\hline
&&&&\multicolumn{4}{c}{Test} &&\multicolumn{3}{c}{Prediction}\\
\cline{5-8} \cline{10-12}
Classifier  &&Features&&Acc$_{agn}$&Acc$_{psr}$&Acc$_{other}$&Acc$_{overall}$&&agn&psr&other \\
\hline

RF&& 20 features (See Table2 left)&&0.992&0.653&0.488&0.939&&959&133&244\\
ANN&& 20 features (See Table2 left)&&0.998&0.673&0&0.917&&1216&120&0\\
RF&& 8 features (See Table2 right)&&0.995&0.633&0.349&0.933&&1112&106&118\\
ANN&& 8 features (See Table2 right)&&1&0.286&0&0.893&&1221&115&0\\
\hline
\end{tabular}}\\
{Note: Column 1: Classification methods.
Column 2: The used features.
Columns 3-6: Test results.
Columns 7-9: Prediction reasults.
}
\end{table*}

\begin{table*}
\centering
\caption{The influence of sample proportion on learning results}\label{Tab9}
\resizebox{\textwidth}{!}{
\begin{tabular}{cccccccccc} 
\hline\hline
&Sample ratio&&\multicolumn{3}{c}{RF} &&\multicolumn{3}{c}{ANN}\\
\cline{4-6} \cline{8-10}
 Sample method &$(n_{non-AGN}/n_{AGN})$ &&Sensitivity&Specificity&Accuracy&&Sensitivity&Specificity&Accuracy\\
\hline
No&0.127&&0.891&0.946&0.939&&0.859&0.956&0.944\\
Undersampling&0.200&&0.891&0.935&0.930&&0.913&0.899&0.901\\
Undersampling&0.500&&0.934&0.869&0.878&&0.946&0.639&0.678\\
Undersampling&1.000&&0.946&0.795&0.814&&0.967&0.398&0.470\\
Oversampling&0.200&&0.891&0.948&0.941&&0.913&0.917&0.916\\
Oversampling&0.500&&0.913&0.937&0.934&&0.946&0.655&0.692\\
Oversampling&1.000&&0.913&0.332&0.930&&0.957&0.356&0.432\\
SMOTE&0.254&&0.891&0.935&0.930&&0.913&0.874&0.879\\
SMOTE&0.508&&0.913&0.912&0.912&&0.946&0.641&0.680\\
SMOTE&1.017&&0.913&0.883&0.887&&0.978&0.339&0.420\\
\hline
\end{tabular}}\\
{Note: Column 1: Sample methods.
Column 2: The sampling ratio, that is, the ratio of the number of non-AGNs to the number of AGNs.
Columns 3 - 5: The performance of ANN classifier in sampling test.
Columns 6 - 8: The performance of RF classifier in sampling test.
}
\end{table*}

We have adopted a step-by-step classification strategy to reduce the large gap in the sample size. However, there is still a class imbalance issue even in the classification process, especially for the first stage of the AGN selection. Undersampling and oversampling are important statistical methods to solve the imbalance of samples. The SMOTE algorithm \citep{SMOTE} is a method to improve the oversampling by constructing new samples; this can reduce the over-fitting consequences of oversampling to some extent. We have studied the effect of different sample proportions on the results (see Table \ref{Tab9}) by different sampling methods (oversampling, undersampling, and SMOTE). There are several important observations for these results. Firstly, in the optimal classifier model used in this paper, the use of a sampling method reduces the accuracy of the classifier. Secondly, in comparison with the ANN algorithm, the RF algorithm has better performance against sample change. Thirdly, in the oversampling, the sensitivity of the non-AGN samples does not increase after the increase of non-AGN samples, which may be due to the over-fitting. However, over-fitting has been avoided in the SMOTE method, which we plan to consider in the future work.

There are some differences between these classifiers' results shown in the preceding section. These results should be treated with caution. Similarly, the accuracy of the second step is not as satisfactory as that of the first one, mainly because the uniqueness of the various sources is not significant enough due to the limited sample size. This also applies to the first step; although we try to expand the non-AGN sample, the gap is still too large. However, as the number of observations in catalogs progress, the situation can be gradually improved.

A potential drawback of this work is that the results are only obtained from the data of the 4FGL catalog. Due to the limitations of astronomical observations, the limited sample used to diagnose the classification of the 4FGL catalog's unassociated samples cannot be completely accurate, and the same applies to our results. In addition, the threshold value of feature selection and the details in the algorithm (random seed, etc.) can directly influence our results, which needs to be further addressed in the future and is beyond the scope of this work.
In addition, the impact of sample selection bias is only discussed and  but not resolved, which needs to further addressed in future work.

\section*{Acknowledgements}
We thank the anonymous referee for very constructive and helpful comments and suggestions, which greatly helped us to improve our paper.
This work is  partially supported by the National Natural Science Foundation of China (Grant Nos. 11763005 and 11873043),
the Science and Technology Foundation of Guizhou Province (QKHJC[2019]1290),
and the Research Foundation for Scientific Elitists of the Department of Education of Guizhou Province (QJHKYZ[2018]068).

\bibliographystyle{raa}
\bibliography{ms_RAA-2020-0191-R2}

\end{document}